\documentclass[aps,pre,twocolumn,amsmath,amssymb,groupedaddress]{revtex4-1}

\usepackage[table]{xcolor}
\definecolor{lightgray}{gray}{0.8}
\definecolor{verylightgray}{gray}{0.9}

\usepackage{graphicx}

\usepackage{array}

\usepackage{bbm}
\usepackage{color}
\usepackage{bm}

\begin{document}

\title{Assessing student's achievement gap between ethnic groups in Brazil}

\author{Luis E C Rocha}\email{luis.rocha@sociology.su.se}
\affiliation{Department of Sociology, Stockholm University, Stockholm, Sweden \\ Department of Public Health Sciences, Karolinska Institutet, Stockholm, Sweden}
\author{Luana de F Nascimento}
\affiliation{Radiation Protection Dosimetry and Calibration group, Belgian Nuclear Research Center, Mol, Belgium}

\date{\today}
\begin{abstract}
Achievement gaps refer to the difference in the performance on examinations of students belonging to different social groups. Achievement gaps between ethnic groups have been observed in several countries with heterogeneous populations. In this paper, we analyze achievement gaps between ethnic populations in Brazil by studying the performance of a large cohort of senior high-school students in a standardized national exam. We separate ethnic groups into the Brazilian states to remove potential biases associated to infrastructure and financial resources, cultural background and ethnic clustering. We focus on the disciplines of mathematics and writing that involve different cognitive functions. We estimate the gaps and their statistical significance through the Welch's t-test and study key socio-economic variables that may explain the existence or absence of gaps. We identify that gaps between ethnic groups are either statistically insignificant ($p<.01$) or small ($2\%$-$6\%$) if statistically significant, for students living in households with low income. Increasing gaps however may be observed for higher income. On the other hand, while higher parental education is associated to higher performance, it may either increase, decrease or maintain the gaps between White and Black, and between White and Pardo students. Our results support that socio-economic variables have major impact on student's performance in both mathematics and writing examinations irrespectively of ethnic backgrounds, giving evidence that genetic factors have little or no effect on ethnic group performance when students are exposed to similar cultural and financial contexts.
\end{abstract}

\maketitle

\section{Introduction}

Education is the systematic process of sharing knowledge and developing skills among people~\cite{Bartlett2007}. It is generally seen as a fundamental right of citizens~\cite{UN1966} but also as an asset both individually and for the country~\cite{Birdsall1999}. Nevertheless, funding is typically limited and several countries struggle to improve their formal educational systems and provide quality education~\cite{Baker2016}. In several cases, strategies are sometimes absent or not well-defined and the already scarce resources end up sub-optimally allocated. A major challenge is to provide similar education opportunities to everyone, given the heterogeneity of students~\cite{Sleeter2008} and resources available at different locations. Moreover, even if biologically different, people are also affected in daily life by the socio-economic context in which they live and thus students have their own individual struggles shaping their learning process.

Examinations are commonly used to assess learning. Scores are then used to rank individuals and to compare the performance of groups of students, such as entire schools, cities, countries, income, ethnicities, sex or other demographic groups. The achievement gap is a term generally used to quantify the persistent difference in the performance of students belonging to different groups in standardized tests~\cite{Ansell2011}. Such gap analysis should not be used to stigmatize groups but to better understand needs and to develop target interventions and policies, aiming to improve education of group members and consequently homogenize performance of the entire population to higher levels. The achievement gap is in the political and academic agenda. Several experiences have shown that it is possible to homogenize student's performance irrespective of their background~\cite{Clark2014}, however, gaps between various groups remain prevalent worldwide~\cite{Carnoy2013a}. In particular, much attention is given to achievement gaps between male and female, low- and high-income, and white and non-white students~\cite{Strand2013, Carnoy2013b}.

Brazil, as several other countries particularly in the Americas, has a relatively recent history of colonization involving spontaneous (in particular, European and Asian) and forced migration (through slavery), and tentative integration to indigenous populations. Although mixing between ethnic groups is generally perceived as higher than in some other countries with similar migration patterns, several indicators suggest a level of segregation by ethnicity, particularly, of so-called whites (typically of European ancestry), blacks (typically descendants of enslaved populations from sub-Saharan Africa between the mid-15th and end-19th century) and mixed-ethnicities (typically those with black and white ancestry, but also white and indigenous, among other combinations), when it comes to income, education, health, empowerment and job opportunities~\cite{Cacciamali2005, Ribeiro2006, Barata2007, Goncalves2014, Augusto2015, Chiavegatto2014}. This country-wide segregation leads to stigma since poverty, criminality, performance or cognitive capacity are associated to particular groups. Aiming to reduce educational gaps of students with diverse income and ethnic backgrounds, governmental affirmative actions have been implemented at higher education in the last decade~\cite{Lloyd2015, Schwartzman2016} but relatively few efforts have been done at the primary and secondary education levels~\cite{Soares2006}.

In this paper, we analyze achievement gaps between ethnic populations in Brazil by studying the performance of a large cohort of senior high-school students in a standardized exam. Education in Brazil is mostly publicly funded and free, where $78.5\%$ of the primary and $70.8\%$ of the secondary schools are managed by the state. Primary education is mostly funded and controlled by the local municipality whereas secondary education is mostly funded and controlled by the states (Brazil is a federative republic)~\cite{Censo2017}. We stratify ethnic groups into the different states to remove potential geographic biases associated to the availability of infrastructure and resources, cultural background or ethnic clustering. We focus our analysis on mathematics and writing that are relevant disciplines involving different cognitive functions. We estimate the gaps and their statistical significance through the advanced Welch's analysis of variance and investigate key socio-economic factors that may explain the existence or absence of gaps in the various cases. We identify that gaps between ethnic groups are positively dependent on the household income and on the parental education level irrespectively of the ethnic group or geographic location. In general, ethnic achievement gaps are statistically insignificant for lower household income and lower parental educational level. However, if gaps are statistically significant at these levels, they are minimal. Our results provide evidence that students exposed to similar low socio-economic conditions have similar performance irrespective of their ethnicity whereas the gaps between ethnic groups sometimes increase for higher income and higher parental education levels.

\section{Materials and methods}

\subsection{Data}

The data corresponds to the performance scores of students in the Brazilian national high school exam (abbreviated ``ENEM'' from the Portuguese ``Exame Nacional do Ensino M\'edio''). This is a standardized national exam, developed and coordinated by the Federal Ministry of Education of Brazil. Since its first edition in 1998, the national exam occurs every year. Although non-mandatory, the individual scores can be currently used by students to compete for enrollment at federal universities (tuition-free) and for other federal high-education scholarship programs to study in private universities (with tuition). Any person at any age who has completed or is about to complete secondary education (the high school) can participate in the exam. The government also uses the results to evaluate the quality of education country-wide. Datasets containing information for each year are publicly available and can be downloaded from ``http://portal.inep.gov.br/microdados''.

The exam is divided in two stages taking place at two different days approximately by the end of the Brazilian academic year, that is late October or early November. The first stage lasts a maximum of four and a half hours and contains 45 multiple choice questions of Natural Sciences (i.e.\ Biology, Physics and Chemistry) and 45 multiple choice questions of Social Sciences (i.e.\ History, Geography, Philosophy and Sociology). The second stage lasts a maximum of five and a half hours and contains 45 multiple choice questions of Languages (i.e.\ Portuguese grammar, Literature, Spanish or English, Arts, Physical Education, and Information and Communication Technologies), 45 multiple choice questions of Mathematics, and 1 written essay (hereafter Writing exam) with a surprise topic. Each exam has four versions that are randomly distributed among the students to avoid cheating but the written essay has a common topic. The multiple choice question scoring is based on the Item response theory that takes into account the level of difficulty of each question and the response patterns of students, giving lower weight to potential guessing~\cite{Hambleton1991, inep2010}. On the other hand, the written essay is graded (and scores are then averaged) by two independent examiners following five criteria of competences related to text interpretation, organization of ideas, grammar, topic, ethics and so on~\cite{inep2017}.

\subsection{Inclusion criteria}

We use data from the exam that occurred on the 8th and 9th of November, 2014. Given the diversity of demographic and socio-economic variables, we define the inclusion criteria aiming to homogenize the sample and avoid spurious influences of external factors. Therefore, the student must (i) have self-declared ethnicity; (ii) be born and raised in Brazil; (iii) be completing high-school in 2014 and be 17 (seventeen) years old (this is the expected age to complete high-school studies); (iv) not be disabled; (v) not be pregnant or lactating ; (vi) not be married, divorced or widowed; (vii) participate in both parts of the exam; (viii) have followed most of the secondary education in publicly administered and funded schools; (ix) live in an urban area; (x) study in an urban school; (xi) live in a house with 6 (six) or less people including the candidate. The variable names (codes) for these inclusion criteria as available in the original data set are specified in the SI.

\subsection{Ethnic structure}

Ethnic and racial classifications are difficult to standardize and generally vary across countries. In Brazil, the standard is to follow the classification made by The Brazilian Institute of Geography and Statistics (abbreviated IBGE in Portuguese www.ibge.gov.br), that is a federal institution responsible for the national census and other demographic and socio-economic official surveys of the Brazilian population. The IBGE classifies Brazilians according to self-declared ``skin-color'' as ``Branco'' (White, mostly of European ancestry but also Middle Easters), ``Negro'' (Black, typically of sub-Saharan African ancestry), ``Pardo'' (that has broad meaning and includes miscegenation of white and black people, white and indigenous, and black and indigenous), ``Amarelo'' (Yellow, from East Asia, particularly Japan, Korea and China) and ``Indigena'' (Indigenous people). We will use the literal translation from now on. In the latest official census before the examination, from 2010, $47.7\%$ of the population self-declared White, $7.6\%$ Black, $43.1\%$ Pardo, $1.1\%$ Yellow, $0.4\%$ Indigenous, and $0.7\%$ none~\cite{Censo2011}. This distribution varies widely across the country. In the national exam ENEM, participants also self-declare as belonging to one of these categories and this classification will be used in our analyses.

\subsection{Statistical Analysis}

To statistically assess if the population mean scores for the different ethnic groups differ, we perform a Welch's Analysis of Variance (ANOVA) test between the means of all ethnic groups. ANOVA splits the aggregate variability found inside a data set into systematic factors, that have statistical influence on the data, and random factors, that have no influence, therefore providing ways to statistically infer the significance of differences in the means. The null hypothesis in our study is that any difference between the ethnic groups is due to chance, i.e.\ $H_0 : x_{\text{white}} = x_{\text{black}} = x_{\text{pardo}} = x_{\text{yellow}} = x_{\text{indigenous}}$. The alternative hypothesis is $H_1 : x_{\text{white}} \neq x_{\text{black}} \neq x_{\text{pardo}} \neq x_{\text{yellow}} \neq x_{\text{indigenous}}$. Since this test only indicates if at least two means are different, we also apply the Welch's t-test for pairs of ethnic groups such that $H_0 : x_{\text{group-1}} = x_{\text{group-2}}$ and $H_1 : x_{\text{group-1}} \neq x_{\text{group-2}}$. The F-statistic for the Welch's t-test is given by

\begin{equation}
F = \dfrac{ \dfrac{1}{k-1} \sum_{j=1}^k w_j(\bar{x}_j - \bar{x}')^2} {1+ \dfrac{2(k-2)}{k^2-1} \sum_{j=1}^k \left(\dfrac{1}{n_j-1}\right)\left(1-\dfrac{w_j}{w}\right)^2},
\end{equation}
where $k$ is the number of ethnic categories, $n_j$ is the size of each category and
\begin{equation}
\bar{x}_j = \dfrac{1}{n_j}\sum_{i=1}^{n_j} x_i, \qquad
s_j^2 = \dfrac{1}{n_j-1}\sum_{i=1}^{n_j} (x_i - \bar{x}_j)^2, \qquad
\end{equation}
\begin{equation}
w_j=\dfrac{n_j}{s_j^2}, \qquad
w = \sum_{j=1}^k w_j, \qquad
\bar{x}' = \dfrac{1}{w}\sum_{j=1}^k w_j \bar{x}_j,
\end{equation}
that is, $\bar{x}_j$ and $s_j^2$ are respectively the mean and variance in each ethnic category $j$. Therefore
\begin{equation}
F \sim F(k-1,df),
\end{equation}
where the degrees of freedom are given by
\begin{equation}
df = \dfrac{k^2-1}{3\sum_{j=1}^k \left(\dfrac{1}{n_j-1}\right)\left(1-\dfrac{w_j}{w}\right)^2}.
\end{equation}

The Welch's statistical test is more robust and recommended if samples have different sizes and variances as the case of our data, but the test also performs well otherwise~\cite{Welch1947, Welch1951, Ruxton2006}. We choose a strong level of confidence $\alpha=0.99$. Therefore, if the $p$-value $<.01$ we reject the null hypothesis and conclude that there is a significant difference between the mean scores. On the other hand, $p$-values $\geqslant .01$ indicate that the null hypothesis is true, i.e.\ mean scores are equal. The achievement gap is given by $\Delta = (\bar{x}_{\text{group-1}} - \bar{x}_{\text{group-2}})/\bar{x}_{\text{group-2}}$, that is, the percentage of the difference between two ethnic groups.

\section{Results}

\subsection{Household income level}

\begin{figure*}[!ht]
\includegraphics[scale=1.0]{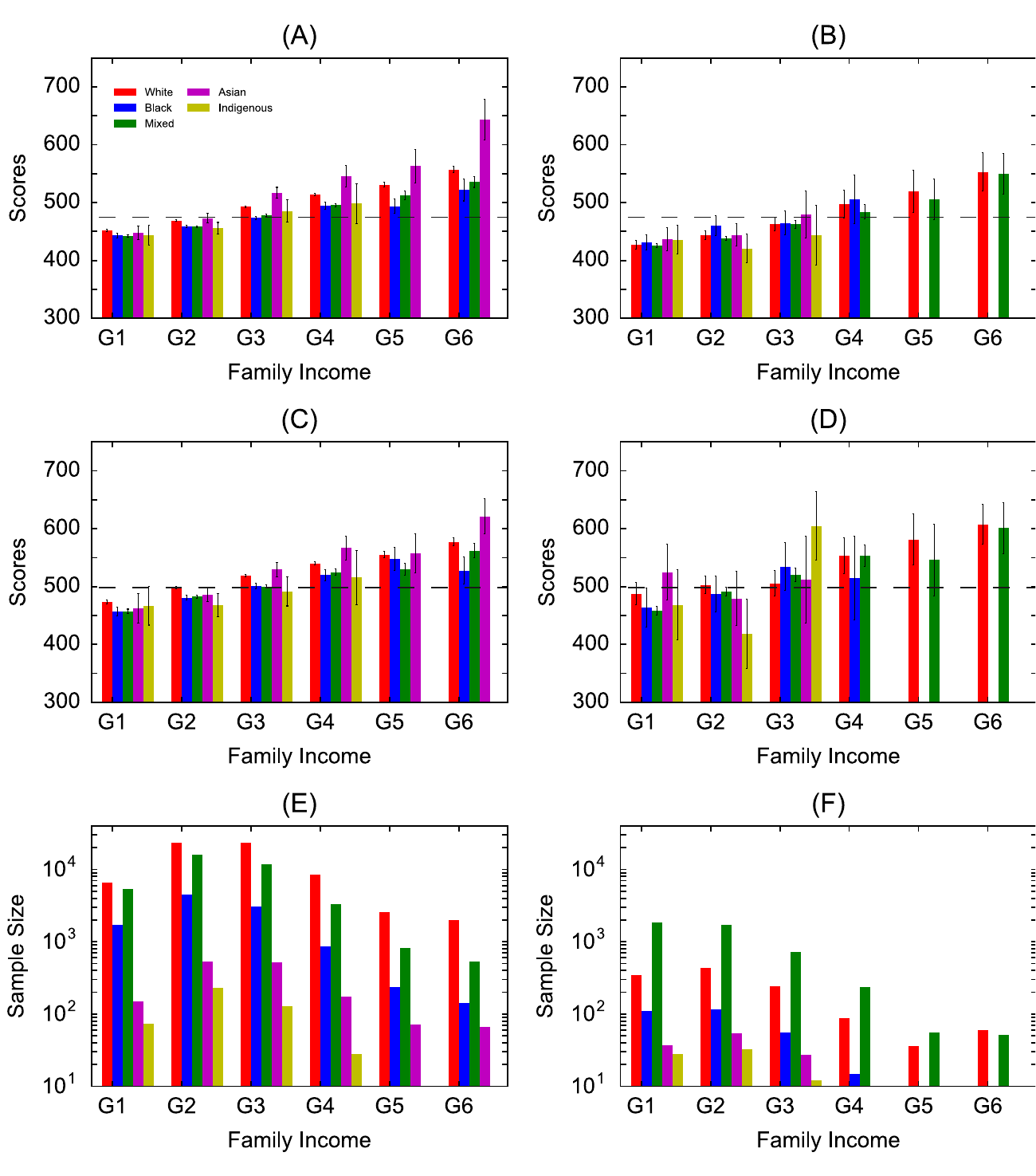}
\caption{{\bf Mean performance score vs. household income level.}
The mean score for 5 ethnic categories and 6 income levels: (A) mathematics in Sao Paulo; (B) mathematics in Amazonas; (C) writing in Sao Paulo; and (D) writing in Amazonas. The sample size for 5 ethnic categories and 6 income levels: (E) Sao Paulo and (F) Amazonas. The household income levels correspond to: G1: up to 1 minimum salary (411 USD in 2014); G2: between 1 and 2; G3: between 2 and 4; G4: between 4 and 6; G5: between 6 and 8; and G6: above 8. Black vertical bars represent the confidence intervals and the horizontal dashed lines are the national mean in each discipline. }
\label{fig:01}
\end{figure*}

Brazil has 23 federal state plus the federal district where the capital of the country is located. Given that primary and secondary education are mostly financed by the local authorities and sociocultural factors and values are also generally defined locally, we study each federal state independently. We first select the states of Sao Paulo and Amazonas because of their contrasting geographic, economic and ethnic variables. Sao Paulo had a population of 41.262,199 people with a White majority in 2010~\cite{Censo2011}, and household income per capita of 1,432 BRL (approx.\ 814 USD) in 2014~\cite{IBGE2015}. Amazonas on the other hand had a population of 3.483,985 people with a Pardo majority in 2010~\cite{Censo2011} and household income per capita of 739 BRL (approx.\ 420 USD) in 2014~\cite{IBGE2015}. For each state, we group students in 5 categories according to their ethnic background and 6 categories according to their family or household income level (G1 to G6, see caption Fig~\ref{fig:01}).

There is generally a trend of increasing mean scores in both mathematics and writing for increasing income in both states (Fig.~\ref{fig:01}A-D). The trend is less clear for writing in Amazonas (Fig.~\ref{fig:01}D). This is likely due to the relatively small populations in some categories in this state (Fig.~\ref{fig:01}F), that also generate relatively larger variance (black lines). We do not use categories with sample sizes smaller than 10 students. Results for Sao Paulo have less variations of the mean scores and smaller confidence intervals, likely due to the relatively larger sample sizes. In general, the two groups with lowest family income (G1 and G2) have mean scores below the national mean, i.e.\ considering all states (dashed lines), whereas the two groups with highest household income have mean scores above the national mean. There is also an apparent increasing gap between some ethnic groups as the income increases. For low household income, we observe mixed results with slightly higher scores for one or another ethnic group depending on the discipline and state. In Sao Paulo there is a clear positive trend of increasing gap for increasing income between Yellow and White students in comparison to Blacks and Pardos. In the state of Amazonas, some ethnic groups are not representative in high income categories but we generally observe similar mean scores for White and Pardo students, i.e.\ absence of gaps, associated to large confidence intervals.

\begin{table*}[ht]
\rowcolors{1}{verylightgray}{}
\setlength\tabcolsep{2.5pt}
\centering
\caption{
{\bf Difference in the mean performance scores and $p$-values for different household income levels.}}
\begin{tabular}{ccccccccccccccc}
\rowcolor{verylightgray}
\hline
& \multicolumn{2}{c}{G1} & \multicolumn{2}{c}{G2} & \multicolumn{2}{c}{G3} & \multicolumn{2}{c}{G4} & \multicolumn{2}{c}{G5} & \multicolumn{2}{c}{G6}       \\
\rowcolor{verylightgray}
        & $\Delta (\%)$ & p-value & $\Delta (\%)$ & p-value & $\Delta (\%)$ & p-value & $\Delta (\%)$ & p-value & $\Delta (\%)$ & p-value & $\Delta (\%)$ & p-value \\ \hline \hline
        
\multicolumn{13}{c}{\bf Sao Paulo - Mathematics}\\ \hline
All    & \multicolumn{2}{c}{$<.01$} & \multicolumn{2}{c}{$<.01$} & \multicolumn{2}{c}{$<.01$} & \multicolumn{2}{c}{$<.01$} & \multicolumn{2}{c}{$<.01$} & \multicolumn{2}{c}{$<.01$} \\
W-B & 2.0 &\cellcolor{lightgray} $<.01$ & 2.2 &\cellcolor{lightgray} $<.01$ & 4.0 &\cellcolor{lightgray} $<.01$ & 3.6 &\cellcolor{lightgray} $<.01$ & 7.3 &\cellcolor{lightgray} $<.01$   & 6.5 &\cellcolor{lightgray} $<.01$    \\
W-P & 2.1 &\cellcolor{lightgray} $<.01$ & 2.3 &\cellcolor{lightgray} $<.01$ & 3.0 &\cellcolor{lightgray} $<.01$ & 3.6 &\cellcolor{lightgray} $<.01$ & 3.5 &\cellcolor{lightgray} $<.01$   & 3.9 &\cellcolor{lightgray} $<.01$    \\
W-Y & 1.0 & .43       & -0.9 & .32       & -4.8 &\cellcolor{lightgray} $<.01$ & -6.0 &\cellcolor{lightgray} $<.01$ & -5.9 & .03     & -14.5 &\cellcolor{lightgray} $<.01$ \\
W-I  & 1.8 & .34       & 2.9 & .01       & 1.6 & .44        & 3.0 & .36      & - & -                    & -        & -             \\ \hline

\multicolumn{13}{c}{\bf Sao Paulo - Writing}\\ \hline
All    & \multicolumn{2}{c}{$<.01$} & \multicolumn{2}{c}{$<.01$} & \multicolumn{2}{c}{$<.01$} & \multicolumn{2}{c}{$<.01$} & \multicolumn{2}{c}{$<.01$} & \multicolumn{2}{c}{$<.01$} \\
W-B & 3.4 &\cellcolor{lightgray} $<.01$ & 3.5 &\cellcolor{lightgray} $<.01$ & 3.6 &\cellcolor{lightgray} $<.01$ & 3.8 &\cellcolor{lightgray} $<.01$ & 1.3 & .50        & 9.0 &\cellcolor{lightgray} $<.01$   \\
W-P & 3.4 &\cellcolor{lightgray} $<.01$ & 3.1 &\cellcolor{lightgray} $<.01$ & 3.8 &\cellcolor{lightgray} $<.01$ & 2.8 &\cellcolor{lightgray} $<.01$ & 4.7 &\cellcolor{lightgray} $<.01$ & 2.6 & .04          \\
W-Y & 2.2 & .42       & 2.3 & .06       & -2.0&  .10      & -4.8 & .01      & -0.3 & .91      & -7.3 &\cellcolor{lightgray} $<.01$   \\
W-I  & 1.3 & .72       & 6.2 & \cellcolor{lightgray} $<.01$ & 5.4 & .04       & 4.6 & .30       & - & -                  & -      & -              \\ \hline

\multicolumn{13}{c}{\bf Amazonas - Mathematics}\\ \hline
All    & \multicolumn{2}{c}{.67}  & \multicolumn{2}{c}{.04} & \multicolumn{2}{c}{.85} & \multicolumn{2}{c}{.41} & \multicolumn{2}{c}{-} & \multicolumn{2}{c}{-}  \\
W-B & -1.0 & .57 & -3.8 & .07 & -0.4 & .88 & -1.7 & .72 & -     & -      & -     & -         \\
W-P & .3  & .79 & 1.3  & .20 & .1   & .96 & 2.8  & .32 & 2.7 & .59 & .6 & .89   \\
W-Y & -2.2 & .38 & -0.2 & .95 & -3.5 & .43 & -      & -       & -     & -      & -     & -         \\
W-I  & -2.0 & .50 & 5.3  & .08 & 4.1  & .45  &-       & -       & -     & -      & -     & -         \\ \hline

\multicolumn{13}{c}{\bf Amazonas - Writing}\\ \hline
All    & \multicolumn{2}{c}{$<.01$}  & \multicolumn{2}{c}{.07} & \multicolumn{2}{c}{.03} & \multicolumn{2}{c}{.54} & \multicolumn{2}{c}{-} & \multicolumn{2}{c}{-} \\
W-B & 5.0 & .22       & 3.2 & .36    & -5.5 & .23      & 7.3 & .31 & - & -           & - & -          \\
W-P & 6.2 & \cellcolor{lightgray} $<.01$ & 2.4 & .16    & -2.8 & .27      & 0 & .99    & 6.3 & .36 & 1.0 & .83 \\
W-Y & -7.3 & .15      & 4.8 & .33    & -1.2 & .87      & - & -           & - & -          & - & -           \\
W-I  & 4.0 & .54       & 18.4 &\cellcolor{lightgray} $<.01$ & -17.9 &\cellcolor{lightgray} $<.01$ & - & -    & - & -          & - & -          \\ \hline

\end{tabular}
\\ W: White; B: Black; P: Pardo; Y: Yellow; I: Indigenous; All: All ethnic categories together. Dark gray highlights $p<.01$.
\label{tab:01}
\end{table*}

The analysis of all ethnic groups together indicates that in the state of Sao Paulo, there are statistically significant ($p$-value $<.01$) differences in the mean scores among groups whereas no differences are observed in the state of Amazonas ($p$-value $\geqslant.01$) (Table~\ref{tab:01}). Results are generally similar for both disciplines. For a more detailed analysis, we now look at pairs of ethnic groups taking the White students as the reference group. The $p$-values indicate statistically significant gaps ($\Delta$) in the mean scores of White (higher mean score) and Black (lower), and White (higher) and Pardo (lower) students for almost all income levels in both disciplines in the state of Sao Paulo. For low income, this gap is slightly higher for writing ($\Delta = 3.4\%$) in comparison to mathematics ($\Delta \sim 2\%$). There is also an increasing gap for higher income levels for mathematics but this increase is not evident for writing. White students tend to perform worse than the Yellow in mathematics when statistically significant gaps are observed, with a substantial gap increase for higher income in favor of Yellow students. In the case of White and Indigenous students, there are typically no statistically significant gaps in mean scores. In the state of Amazonas, we observe no statistical significant gaps with the exception of writing scores for White and Indigenous students. However, the unusual large gap ($\Delta \sim 18.4\%$ to $\Delta = -17.9\%$) from favoring White to favoring Indigenous students between two subsequent income levels suggest outliers and would need more data for a careful analysis. Looking at the mean scores, we generally identify worse performance ($\Delta < 0$) of White students in comparison to Black, Pardo and Yellow students, particularly in the case of mathematics.

Analysing the mean scores of all Brazilian federal states independently, there are no statistically significant gaps between ethnic groups in most of the scenarios for very low and very high income households (See Tables I and II in SI). Although statistically significant results for very low income in three states (MG, RJ and SP) indicate better performance of White students in comparison to Blacks and Pardos in mathematics, results from one state (MG) also indicate better performance of Whites in comparison to Yellows. For writing, in a few states White students also showed better performance in comparison to Black, Pardo and Yellow students (with gaps slightly larger than in the case of mathematics) but there are generally no strong trends favoring one or another ethnicity.

\subsection{Parental education level}

\begin{figure*}[!ht]
\includegraphics[scale=1.0]{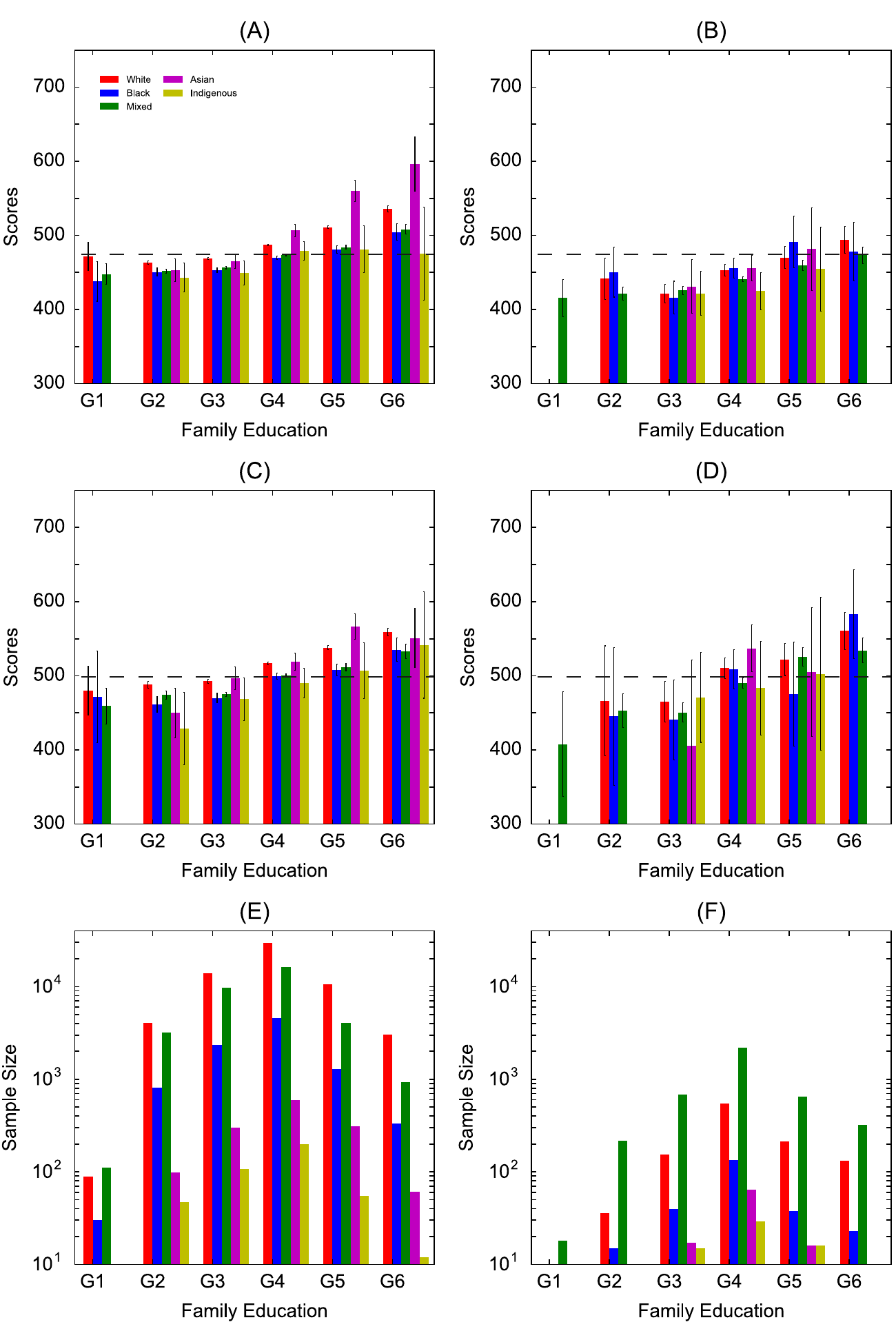}
\caption{{\bf Mean performance score vs. parental education.}
The mean score for 5 ethnic categories and 6 education levels: (A) mathematics in Sao Paulo; (B) mathematics in Amazonas; (C) writing in Sao Paulo; and (D) writing in Amazonas. The sample size for 5 ethnic categories and 6 parental education levels: (E) Sao Paulo and (F) Amazonas. The education levels correspond to: G1: both parents have no formal education; G2: at least one parent has completed 4 years of formal education; G3: at least one parent has completed 8 years of formal education (i.e.\ completed primary education) or completed primary but not secondary education (i.e.\ started but not completed); G4: at least one parent has completed secondary education or has completed secondary but not tertiary education (i.e.\ started but not completed); G5: at least one parent has completed tertiary education; and G6: at least one parent has completed post-graduate studies. Black vertical bars represent the confidence intervals and the horizontal dashed lines are the national mean in each discipline.}
\label{fig:02}
\end{figure*}

In this section, we look at the potential gaps between ethnic groups and the education level of the student's parents. Once again, for each state, we group students in 5 categories according to their ethnic background and in 6 categories according to their parents education (G1 to G6, see caption Fig~\ref{fig:02}). We observe a general trend of increasing mean scores for students with more educated parents but mean scores do not differ much for groups of less educated parents (G1 to G3) (Fig~\ref{fig:02}A-D). This happens for both disciplines in both states but in Amazonas the confidence intervals are generally larger (Fig~\ref{fig:02}B,D). In Sao Paulo, White students tend to perform better than other ethnic groups with the exception that Yellow students perform substantially better in mathematics in case of highly educated parents (Fig~\ref{fig:02}A) and also in group G5 in the case of writing (Fig~\ref{fig:02}C). In the state of Amazonas, there is a mixed pattern in which the best performances alternate between ethnic groups, including for highly educated parents, for both disciplines (Fig~\ref{fig:02}B,D). Note also that the distribution of the sample sizes are higher in mid-level education (G4) in both states (Fig~\ref{fig:02}E-F) whereas we observe larger samples sizes in lower household income levels in Fig~\ref{fig:01}E,F.

\begin{table*}[ht]
\rowcolors{1}{verylightgray}{}
\setlength\tabcolsep{2.5pt}
\centering
\caption{
{\bf Difference in the mean performance scores and $p$-values for different parental education levels.}}
\begin{tabular}{ccccccccccccc}
\rowcolor{verylightgray}
\hline
        & \multicolumn{2}{c}{G1} & \multicolumn{2}{c}{G2} & \multicolumn{2}{c}{G3} & \multicolumn{2}{c}{G4} & \multicolumn{2}{c}{G5} & \multicolumn{2}{c}{G6}  \\
\rowcolor{verylightgray}
        & $\Delta (\%)$ & p-value & $\Delta (\%)$ & p-value & $\Delta (\%)$ & p-value & $\Delta (\%)$ & p-value & $\Delta (\%)$ & p-value & $\Delta (\%)$ & p-value \\ \hline \hline

\multicolumn{13}{c}{\bf Sao Paulo - Mathematics}\\ \hline
All    & \multicolumn{2}{c}{.07} & \multicolumn{2}{c}{$<.01$} & \multicolumn{2}{c}{$<.01$} & \multicolumn{2}{c}{$<.01$} & \multicolumn{2}{c}{$<.01$} & \multicolumn{2}{c}{$<.01$} \\
W-B & 7.4 & .04  & 2.8 & \cellcolor{lightgray} $<.01$   & 3.5 & \cellcolor{lightgray} $<.01$      & 3.6    & \cellcolor{lightgray} $<.01$ & 6.1 & \cellcolor{lightgray} $<.01$  & 6.0  & \cellcolor{lightgray} $<.01$ \\
W-P & 5.2 & .05 & 2.5  & \cellcolor{lightgray} $<.01$ & 2.7   & \cellcolor{lightgray} $<.01$      & 2.8    & \cellcolor{lightgray} $<.01$ & 5.4  & \cellcolor{lightgray} $<.01$ & 5.2  & \cellcolor{lightgray} $<.01$ \\
W-Y & -    & -      & 2.3 & .19        & 0.9   & .42           & -4.0 & \cellcolor{lightgray} $<.01$   & -9.1 & \cellcolor{lightgray} $<.01$ & -10.7 & \cellcolor{lightgray} $<.01$ \\
W-I  & -    & -      & 4.5  & .04        & 4.3  & .02            & 1.7    & .21       & 6.0 & .06        & 11.9 & .06 \\ \hline

\multicolumn{13}{c}{\bf Sao Paulo - Writing}\\ \hline
All    & \multicolumn{2}{c}{0.60} & \multicolumn{2}{c}{$<.01$} & \multicolumn{2}{c}{$<.01$} & \multicolumn{2}{c}{$<.01$} & \multicolumn{2}{c}{$<.01$} & \multicolumn{2}{c}{$<.01$} \\
W-B & 1.8 & .81 & 5.7 & \cellcolor{lightgray} $<.01$ & 4.7 & \cellcolor{lightgray} $<.01$ & 3.4  & \cellcolor{lightgray} $<.01$ & 5.7  & \cellcolor{lightgray} $<.01$ & 4.4 & \cellcolor{lightgray} $<.01$\\
W-P & 4.4 & .32 & 3.0 & \cellcolor{lightgray} $<.01$ & 3.6 & \cellcolor{lightgray} $<.01$ & 3.3  & \cellcolor{lightgray} $<.01$ & 5.0  & \cellcolor{lightgray} $<.01$ & 4.8 & \cellcolor{lightgray} $<.01$ \\
W-Y & -     & -    & 8.2 & .03       & -0.9 &  .56      & -0.4 & .72     & -5.2 & \cellcolor{lightgray} $<.01$  & 1.5 & .68 \\
W-I  & -      & -    & 12.9 & .02     & 5.0 & .10       & 5.3  & \cellcolor{lightgray} $<.01$ & 5.9  & .10       & 3.2 & .60\\ \hline

\multicolumn{13}{c}{\bf Amazonas - Mathematics}\\ \hline
All    & \multicolumn{2}{c}{-}  & \multicolumn{2}{c}{0.12} & \multicolumn{2}{c}{0.89} & \multicolumn{2}{c}{$<.01$} & \multicolumn{2}{c}{0.31} & \multicolumn{2}{c}{0.16} \\
W-B & - & - & -2.0 & .67  & 1.3 & .68  & -0.7  & .69        & -4.4 & .27 & 3.3 & .45 \\
W-P & - & - & 4.6  & .18 & -1.0 & .55   & 2.7       & \cellcolor{lightgray} $<.01$  & 2.3  & .2  & 4.4 & .05 \\
W-Y & - & - & - & -  & -2.3 & .60  & -0.6       & .77        & -2.4 & .68 & -     & - \\
W-I  & - & - & -  & -  & -0.1  & .98  & 6.4        & .03          & 3.4 & .58  & -     & - \\ \hline

\multicolumn{13}{c}{\bf Amazonas - Writing}\\ \hline
All    & \multicolumn{2}{c}{-}  & \multicolumn{2}{c}{.93} & \multicolumn{2}{c}{0.70} & \multicolumn{2}{c}{$<.01$} & \multicolumn{2}{c}{0.66} & \multicolumn{2}{c}{0.10}  \\
W-B & - & - & 4.6  & .72    & 5.5 & .41      & 0.3   & .90   & 9.4  & .20 & -4.0 & .47 \\
W-P & - & - & 2.8  & .73    & 3.2 & .34      & 4.0    & .01   & -0.7 & .77  & 4.8 & .09 \\
W-Y & - & - & - & -             & -13.7 & .30   & -5.1  & .13   & 3.3 & .69  & - & - \\
W-I  & - & - & - & -             & -1.1 & .87     & 5.4   & .40   & 3.8 & .70  & - & - \\ \hline

\end{tabular}
\\ W: White; B: Black; P: Pardo; Y: Yellow; I: Indigenous; All: All ethnic categories together. Dark gray highlights $p<.01$.
\label{tab:02}
\end{table*}

The analysis of scores altogether indicates that the gap is statistically significant for both disciplines in Sao Paulo but not in the state of Amazonas (Tab~\ref{tab:02}). Looking at pairs of ethnic groups, the gap is statistically significant ($p$-value $<.01$) between White (higher scores) and Black (lower) students, and White (higher) and Pardo (lower) students for all groups but G1. Furthermore, in highly educated families, there is also statistical significance between White and Yellow students, particularly in mathematics, with Yellow students achieving higher mean scores. Generally speaking, no statistically significant gaps are observed between White and Yellow students. In the case of Amazonas, there is no statistical significant differences, with the exception of White and Black in G4 for mathematics. In the lower end (i.e.\ low parental education), the gaps ($\Delta$) between ethnic groups are generally higher in the respective educational category in comparison to the gaps observed in the categories defined by low household income, whereas the opposite occurs in the higher end, that is, relatively speaking the gap is smaller in highly educated families than in families with higher income. In other words, household income levels seem to be a stronger indicator (than parents education) of higher scores. Our statistical analysis supports that in the state of Amazonas, there is no clear trends in the gaps, i.e.\ for a given parental education level, one or another ethnic group presents higher performance in comparison to the other.

\begin{table*}[ht]
\rowcolors{1}{verylightgray}{}
\setlength\tabcolsep{2.5pt}
\centering
\caption{
{\bf The mean performance scores and $p$-values for different household income levels.}}
\begin{tabular}{ccccccccccccc}
\hline

        & \multicolumn{2}{c}{G1} & \multicolumn{2}{c}{G2} & \multicolumn{2}{c}{G3} & \multicolumn{2}{c}{G4} & \multicolumn{2}{c}{G5} & \multicolumn{2}{c}{G6}        \\
\rowcolor{verylightgray}
        & $\Delta (\%)$ & p-value & $\Delta (\%)$ & p-value & $\Delta (\%)$ & p-value & $\Delta (\%)$ & p-value & $\Delta (\%)$ & p-value & $\Delta (\%)$ & p-value \\ \hline \hline

\multicolumn{13}{c}{\bf Sao Paulo - Mathematics}\\ \hline
M-F  & 7.6 &\cellcolor{lightgray} $<.01$ & 7.6 &\cellcolor{lightgray} $<.01$  & 7.9  &\cellcolor{lightgray} $<.01$ & 8.1 &\cellcolor{lightgray} $<.01$ & 8.4 &\cellcolor{lightgray} $<.01$ & 8.4 &\cellcolor{lightgray} $<.01$  \\ \hline

\multicolumn{13}{c}{\bf Sao Paulo - Writing}\\ \hline
M-F  & -5.9 &\cellcolor{lightgray} $<.01$ & -7.4 &\cellcolor{lightgray} $<.01$ & -8.5 &\cellcolor{lightgray} $<.01$ & -8.7 &\cellcolor{lightgray} $<.01$ & -10.4 &\cellcolor{lightgray} $<.01$ & -7.9 &\cellcolor{lightgray} $<.01$  \\ \hline

\multicolumn{13}{c}{\bf Amazonas - Mathematics}\\ \hline
M-F  & 5.7 &\cellcolor{lightgray} $<.01$ & 6.0 &\cellcolor{lightgray} $<.01$ & 7.1 &\cellcolor{lightgray} $<.01$ & 3.5 & .12 & 9.6 & .05 & 8.0 & .07  \\ \hline

\multicolumn{13}{c}{\bf Amazonas - Writing}\\ \hline
M-F  & -4.7 &\cellcolor{lightgray} $<.01$ & -5.1 &\cellcolor{lightgray} $<.01$ & -3.8 & .05 & -6.1 & .03 & -10.4 & .14  & -5.5 & .22 \\ \hline

\end{tabular}
\\ M: Male; F: Female
\label{tab:03}
\end{table*}

\begin{table*}[ht]
\rowcolors{1}{verylightgray}{}
\setlength\tabcolsep{2.5pt}
\centering
\caption{
{\bf The mean performance scores and $p$-value for different parental education levels.}}
\begin{tabular}{ccccccccccccc}
\hline

        & \multicolumn{2}{c}{G1} & \multicolumn{2}{c}{G2} & \multicolumn{2}{c}{G3} & \multicolumn{2}{c}{G4} & \multicolumn{2}{c}{G5} & \multicolumn{2}{c}{G6} \\
\rowcolor{verylightgray}
        & $\Delta (\%)$ & p-value & $\Delta (\%)$ & p-value & $\Delta (\%)$ & p-value & $\Delta (\%)$ & p-value & $\Delta (\%)$ & p-value & $\Delta (\%)$ & p-value \\ \hline \hline

\multicolumn{13}{c}{\bf Sao Paulo - Mathematics}\\ \hline
M-F  & 10.9 &\cellcolor{lightgray} $<.01$ & 8.1 &\cellcolor{lightgray} $<.01$ & 8.4  &\cellcolor{lightgray} $<.01$ & 8.5 &\cellcolor{lightgray} $<.01$ & 8.6 &\cellcolor{lightgray} $<.01$ & 8.5 &\cellcolor{lightgray} $<.01$ \\ \hline

\multicolumn{13}{c}{\bf Sao Paulo - Writing}\\ \hline
M-F  & -7.6 & .08 & -6.2 &\cellcolor{lightgray} $<.01$ & -6.5 &\cellcolor{lightgray} $<.01$ & -7.5 &\cellcolor{lightgray} $<.01$ & -7.7 &\cellcolor{lightgray} $<.01$ & -9.9 &\cellcolor{lightgray} $<.01$ \\ \hline

\multicolumn{13}{c}{\bf Amazonas - Mathematics}\\ \hline
M-F  & 2.9 & .96 & 6.1 &\cellcolor{lightgray} $<.01$ & 5.5 &\cellcolor{lightgray} $<.01$ & 6.4 &\cellcolor{lightgray} $<.01$ & 7.2 &\cellcolor{lightgray} $<.01$ & 7.0 &\cellcolor{lightgray} $<.01$ \\ \hline

\multicolumn{13}{c}{\bf Amazonas - Writing}\\ \hline
M-F  & -15.6 & .33 & -3.6 & .48 & -4.9 & .06 & -3.8 &\cellcolor{lightgray} $<.01$ & -3.6 & .08 & -9.4 &\cellcolor{lightgray} $<.01$ \\ \hline

\end{tabular}
\\ M: Male; F: Female
\label{tab:04}
\end{table*}

The analysis of the mean performance scores of all Brazilian federal states independently indicates that in some states (but not in the majority) the gaps between ethnic groups are statistically significant in both very low and very high levels of parental education (See Tables III and IV in SI). In the case of mathematics, in the two states (MG and SP) that statistically significant differences are observed for all combinations of ethnic groups, one (MG) shows that the gap decreased for higher education level (in comparison to the lower education level) whereas in the other (SP) the gap increased for both White-Black and White-Pardo cases. For the White-Yellow analysis, the gap slightly increased for higher parental education in favor of Whites. However, for higher education levels, the gaps are higher and favor Yellow students in some states (PR and SP) and White students in some others (MG and RJ). The differences are not statistically significant for all cases of Indigenous students but in one case of very low education, in which the gap is relatively large in favor of White students. In the case of writing, either the gap decreased for White-Black and White-Pardo from very low to very high parental education level in MG or remained approximately the same in the case of SP.

\subsection{Students sex}

To provide a baseline for comparison, we now analyze the mean performance scores of students according to their self-declared sex, i.e.\ male or female. In this case, we do not distinguish the ethnic backgrounds but only the sex. We use the same categories as before for both household income and parental education. The results are statistically significant for all categories for both disciplines in the state of Sao Paulo but only for low income categories in the state of Amazonas (Table~\ref{tab:03}). Here we observe relatively larger gaps between male and female students in comparison to gaps observed between ethnic groups in the previous sections (Table~\ref{tab:01} and Table~\ref{tab:02}). In all cases, male students perform better in mathematics whereas female students perform better in writing. There are no strong variations in the gaps between household income levels (approx. $6\%$ to $10\%$).

If we categorize students according to their parental education level, the gaps are generally statistically significant in both states for most levels of education (Table~\ref{tab:04}). Also in this case, male students perform better than females in mathematics and vice-versa for writing. The gaps are also higher in this case than if students are categorized by household income level. This is the same trend observed when studying the ethnic groups. Finally, we note that the gaps are relatively similar for different parental education levels in the state of Sao Paulo whereas they vary more in the state of Amazonas. However in the later case, the gaps are not always statistically significant.

\section{Discussion}

In the early 20th century, several scholars believed that intelligence was mostly due to evolutionary characteristics related to ethnicity (also known as ``race gap''). This theory was widely used to support eugenic laws and immigration policies. In the 1940s and 1950s, researchers started to focus on other potential causes, such as environmental and cultural factors, hypothesizing that reducing economic inequality for instance would homogenize the performance of students in exams and tests~\cite{Jackson2004}. The discovery of the DNA structure in 1953 together with the educational desegregation of the southern United States triggered a new wave of studies correlating intelligence to genetic traits. Similar ideas and beliefs were developed in Brazil~\cite{daSilva2000}. Research has followed, mostly attempting to find a genetic component to the observed gaps in intelligent test scores (I.Q.) as well as its relative contribution to the overall performance of students. However, these studies provided insufficient evidence of genetic traits defining I.Q. performance~\cite{Deary2009}. Recent analyses suggest that I.Q. score gaps between White and Black students for example have been decreasing in the USA~\cite{Dickens2006}. Furthermore, studies comparing I.Q. scores from adopted African-American, Indian and Asian children raised by White parents with scores of children raised by parents of the same ethnic group have shown that gaps disappear in mixed-families~\cite{Thomas2017}. These studies suggest that not only household income but also a combination of socio-economic variables, such as removal of stigma, parental background, affirmative actions~\cite{Scarr1976, Ijzendoorn2005}, and better health and nutrition~\cite{Dani2005} may homogenize intelligence scores. Assessing intelligence through I.Q. scores and identifying the variables driving individual performance in such tests however remain controversial. In recent decades, several researchers have turned attention to broader assessments of intelligence or cognitive abilities by analyzing the performance of students through formal examinations of disciplines taught at primary and secondary schools.

Achievement gaps between ethnic groups as estimated through formal examinations exist in several countries but the gap size is not consistent across countries and groups~\cite{Hanushek2010, Zhang2011,Azzolini2012, Zhao2016}. Our analysis of the Brazilian context shows a substantial variation of gaps within the country without a clear dominant group, although White students tend to perform better. In the case of students living in low income households, there are generally no statistically significant gaps and sometimes only small (typically bellow $6\%$, but varying from $1.7$ to $23.1\%$) statistically significant gaps are observed between the various ethnic groups. For comparison, we have also estimated the performance of males and females (irrespective of their ethnic background) and observed that in our context gaps are statistically significant (similarly to some other countries~\cite{Cloer2001, Hyde1990}), higher than between ethnic groups, and relatively constant for all income levels. These results support that socio-cultural and environmental factors, such as poverty, lack of access to extra-curricular activities, malnutrition and lack of basic infrastructure, related to health or safety~\cite{Weihua2011}, are potential relevant variables that negatively affect the performance of students~\cite{Maluccio2009} irrespectively of their ethnicity, potentially more than intrinsic cognitive factors. If performance, as a group, was mostly dependent on genetics or other biological traits, one would observe consistent gaps between ethnic groups irrespective of income. At best, we may conclude that ``speciation'' was not sufficiently large to distinguish those groups. In the case of males and females, the analysis of data from the 2003 Programme for International Student Assessment (PISA) indicates that the mathematics gap disappears and the reading gap becomes larger in more gender-equal cultures~\cite{Guiso2008}, also ruling out that genetics have strong influence on group performance, at least for mathematics. This of course does not reject the hypothesis that genetics may affect people intellectual abilities individually.

At higher income levels, scores are generally higher. In Brazil, official educational policies began to look at different backgrounds and abilities after the new federal constitution of 1988~\cite{Saviani2005}. Before this constitution, public education focused on preparing the working class to develop technical skills, while the development of higher cognitive abilities such as creativity, arts, interpretation, was essentially available to those on upper classes through extra-curricular activities in private schools~\cite{daSilva2002}. Nevertheless, full migration away from the technical-based education still faces challenges such as lack of formal training of teachers, poor availability of education resources, e.g. computers, Internet or other materials, and conservatism. Therefore, students without personal resources and opportunities for private or non-governmental extra curricular activities will likely continue in disadvantage and expected to perform worse on average than their higher income colleagues. The intellectual environment in which students live also affects their performance~\cite{Marteleto2013}. The educational level of parents have a generally positive impact on the average scores of students of all ethnic groups. Contrastingly, it does not increase the gap between Whites and Blacks, and between Whites and Pardos in writing in Sao Paulo but does decrease these gaps in the state of Minas Gerais (abbreviated MG in SI, where gaps are also statistically significant). For mathematics, the gaps between these groups increase in Sao Paulo but decrease in Minas Gerais (with larger decrease between Whites and Blacks). Altogether, these results provide a complex picture that makes difficult to identify particular factors increasing or decreasing gaps in the different areas of Brazil. However, they further support our previous conclusions that student's performance may homogenize given appropriate material, cultural and peer (e.g.\ family) support, irrespectively of ethnic background.

Our results also indicate increasing achievement gaps between ethnic groups at higher household income levels for Sao Paulo students but these gaps are not statistically significant for other regions. In Sao Paulo, Yellow students score best in mathematics, followed by White, Pardo, Black and then Indigenous students. In writing, Whites score best, followed by Yellows, Blacks and Pardos. Household income affects positively the scores of all groups, particularly in mathematics but also in writing, and larger gaps are observed for higher income levels. It is unclear why the increase in income does not affect equally all ethnic groups in Sao Paulo. The current income is a timely measure and does not reflect the historical influences that steady, or absence of steady (traditional vs. new ``rich'' families), household finances might have in the family social context and overall access of education and culture. Brazilian students of Asian ancestry belonging to higher income groups achieve the best scores for mathematics. This may be related to the cultural background and educational policies. The Asian migration to Brazil started in the early 20th century with Japaneses and then shifted towards Koreans and Chinese in the 1980s and 1990s. All these groups have pronounced heritage identity with strong spatial clustering and sense of community, relatively few inter-ethnic marriages and thus few mixed-descendants~\cite{Saito1973} which likely contribute to sustain the high expectations of student performance in this group. We have observed for example the strong effect of increasing parental education of Yellow students, particularly in mathematics. This cultural homogeneity is not as strongly observed for Blacks and Whites in Brazil as is observed in the USA. On the contrary, the large percentage of ``mixed'' populations in Brazil (i.e.\ the Pardos) linking Whites and Blacks without clear cultural borders result in a highly heterogeneous society where ethnic peer-pressure becomes weak. At the same time, Brazil has an European-centric educational system, where much attention is given to history and culture of Europe, to which White students directly relate and thus feel empowered and motivated, particularly in southern states where European immigration is more recent. Efforts aiming to decrease this bias include recent federal laws (11.645/08 from 2008) obliging teaching of African history and culture, Afro-Brazilian culture and Indigenous history in Brazilian schools. The implementation of such regulations is however relatively slow and faces some resistance from part of the population, still influenced by colonial times when African and Indigenous populations were marginalized and considered of less importance~\cite{Pereira2011}.

\section{Conclusion}

The achievement gap between students of different ethnic backgrounds has been observed in various countries with multi-ethnic populations. Brazil is a representative middle-income country that struggles to provide free quality education to its entire geographically spread population. Our study uses data of a standardized national exam to estimate the gap in mathematics and writing between ethnic groups of students attending public schools at different parts of the country. Using advanced Welch's t-test analysis, we have identified the absence or negligible performance gaps between ethnic groups of students living in low income households. Increasing gaps however were observed in some cases for students living in households with higher income levels. On the other hand, while higher parental education is associated to higher performance, it may either increase, decrease or maintain stable the gaps between White and Black, and between White and Pardo students. Overall, our analysis provides evidence that socio-economic variables play a major role in student's performance in mathematics and writing examinations irrespectively of ethnic backgrounds and gives evidence that genetic factors may have little or no effect on group performance. As such, we recommend the design of tailored policies targeting the improvement of well-being, health, work opportunities, income and family education to vulnerable students in low-income settings irrespectively of their ethnic background. We further recommend target policies of empowerment to Black, Pardo and Indigenous students to reduce stigma and to provide socio-cultural conditions to make them motivated and competitive, aiming to homogenize performance scores to higher levels.



\begin{thebibliography}{0}%
\makeatletter
\providecommand \@ifxundefined [1]{%
 \@ifx{#1\undefined}
}%
\providecommand \@ifnum [1]{%
 \ifnum #1\expandafter \@firstoftwo
 \else \expandafter \@secondoftwo
 \fi
}%
\providecommand \@ifx [1]{%
 \ifx #1\expandafter \@firstoftwo
 \else \expandafter \@secondoftwo
 \fi
}%
\providecommand \natexlab [1]{#1}%
\providecommand \enquote  [1]{``#1''}%
\providecommand \bibnamefont  [1]{#1}%
\providecommand \bibfnamefont [1]{#1}%
\providecommand \citenamefont [1]{#1}%
\providecommand \href@noop [0]{\@secondoftwo}%
\providecommand \href [0]{\begingroup \@sanitize@url \@href}%
\providecommand \@href[1]{\@@startlink{#1}\@@href}%
\providecommand \@@href[1]{\endgroup#1\@@endlink}%
\providecommand \@sanitize@url [0]{\catcode `\\12\catcode `\$12\catcode
  `\&12\catcode `\#12\catcode `\^12\catcode `\_12\catcode `\%12\relax}%
\providecommand \@@startlink[1]{}%
\providecommand \@@endlink[0]{}%
\providecommand \url  [0]{\begingroup\@sanitize@url \@url }%
\providecommand \@url [1]{\endgroup\@href {#1}{\urlprefix }}%
\providecommand \urlprefix  [0]{URL }%
\providecommand \Eprint [0]{\href }%
\providecommand \doibase [0]{http://dx.doi.org/}%
\providecommand \selectlanguage [0]{\@gobble}%
\providecommand \bibinfo  [0]{\@secondoftwo}%
\providecommand \bibfield  [0]{\@secondoftwo}%
\providecommand \translation [1]{[#1]}%
\providecommand \BibitemOpen [0]{}%
\providecommand \bibitemStop [0]{}%
\providecommand \bibitemNoStop [0]{.\EOS\space}%
\providecommand \EOS [0]{\spacefactor3000\relax}%
\providecommand \BibitemShut  [1]{\csname bibitem#1\endcsname}%
\let\auto@bib@innerbib\@empty
\end{thebibliography}%


\begin{thebibliography}{10}

\bibitem{Bartlett2007}
Steve Bartlett and Diana Burton.
\newblock Introduction to Education Studies. 3rd Edition
\newblock Sage Publications Ltd 400p (2010)

\bibitem{UN1966}
UN General Assembly.
\newblock International Covenant on Economic, Social and Cultural Rights.
\newblock United Nations, Treaty Series 993, 3 (1966) 

\bibitem{Birdsall1999}
Nancy Birdsall.
\newblock Education: The people's asset.
\newblock Center on Social and Economic Dynamics - Working paper 5
\newblock Washington: Brookings Institution Press (1999)

\bibitem{Baker2016}
Bruce D. Baker.
\newblock Does money matter in education?
\newblock Washington, DC: The Albert Shanker Institute (2016)

\bibitem{Sleeter2008}
Christine Sleeter.
\newblock An invitation to support diverse students through teacher education.
\newblock Journal of Teacher Education 59:3 212--219 (2008)

\bibitem{Ansell2011}
Susan Ansell.
\newblock Editorial projects in education research center - Issues A-Z: Achievement gap.
\newblock Education Week (July 7, 2011)

\bibitem{Censo2017}
\newblock {Censo escolar da educa\c{c}\~ao b\'asica 2016 -- Notas estat\'isticas}.
\newblock Instituto Nacional de Estudos e Pesquisas Educacionais An\'isio Teixeira (2017)

\bibitem{Censo2011}
\newblock Características da popula\c{c}\~ao e dos domic\'ilios -- Resultados do universo.
\newblock Instituto Brasileiro de Geografia e Estat\'istica, Rio de Janeiro (2011)

\bibitem{IBGE2015}
\newblock IBGE divulga renda domiciliar per capita 2014.
\newblock Instituto Brasileiro de Geografia e Estat\'istica, Rio de Janeiro (2015)

\bibitem{Clark2014}
Julia V Clark.
\newblock Closing the achievement gap from an international perspective.
\newblock Springer: Netherlands 324p (2014)

\bibitem{Carnoy2013a}
Martin Carnoy and Richard Rothstein.
\newblock International tests show achievement gaps in all Countries, with big gains for U.S. disadvantaged students.
\newblock Economic Policy Institute Report (January 15, 2013)

\bibitem{Carnoy2013b}
Martin Carnoy and Richard Rothstein.
\newblock What do international tests really show about U.S. student performance?
\newblock Economic Policy Institute Report (January 28, 2013)

\bibitem{Strand2013}
Steve Strand.
\newblock What accounts for ethnic achievement gaps in secondary schools in England?
\newblock British Educational Research Association 4:1-4 (2013)

\bibitem{Augusto2015}
Nat\'alia Augusto and Jos\'e Eduardo Roselino and Andrea Rodrigues Ferro. 
\newblock A Evolu\c{}c\~ao Recente da Desigualdade entre Negros e Brancos no Mercado de Trabalho das Regi\~oes Metropolitanas do Brasil.
\newblock Revista Pesquisa \& Debate 26:2 105 -- 127 (2015)

\bibitem{Cacciamali2005}
Maria Cristina Cacciamali and Guilherme Issamu Hirata.
\newblock A influ\^encia da ra\c{c}a e do g\^enero nas oportunidades de obten\c{c}\~ao de renda -- uma an\'alise da discrimina\c{c}\~ao em mercados de trabalho distintos: {B}ahia e {S}\~ao Paulo.
\newblock Estudos Econ\^omicos 35:4 (2005)

\bibitem{Ribeiro2006}
Carlos Antonio Costa Ribeiro.
\newblock Class, race, and social mobility in Brazil.
\newblock Dados - Revista de Ci\^encias Sociais 49:4 833 -- 873 (2006)

\bibitem{Barata2007}
Rita Barradas Barata and M\'arcia Furquim de Almeida and Cl\'audia Valencia Montero and Zilda Pereira da Silva.
\newblock Health inequalities based on ethnicity in individuals aged 15 to 64, Brazil, 1998.
\newblock Cadernos Sa\'ude P\'ublica 23(2):305--313 (2007)

\bibitem{Goncalves2014}
Josimar Gon\c{c}alves de Jesus.
\newblock Diferen\c{c}as de rendimento entre negros e brancos no Brasil: Evolu\c{c}\~ao e determinantes.
\newblock Master Thesis. Publisher University of Sao P\~aulo, S\~ao Paulo (2014)

\bibitem{Chiavegatto2014}
Alexandre Dias Porto Chiavegatto Filho and Hiram Beltr\'an S\'anchez and Ichiro Kawachi.
\newblock Racial disparities in life expectancy in Brazil: Challenges from a multiracial society.
\newblock American Journal of Public Health 104(11): 2156--2162 (2014)

\bibitem{Schwartzman2016}
Luisa Farah Schwartzman and Angela Randolpho Paiva.
\newblock Not just racial quotas: Affirmative action in Brazilian higher education 10 years later.
\newblock British Journal of Sociology of Education 37:4 548--566 (2016)

\bibitem{Lloyd2015}
Marion Lloyd.
\newblock A decade of affirmative action in Brazil: Lessons for the global debate.
\newblock Mitigating Inequality: Higher Education Research, Policy, and Practice in an Era of Massification and Stratification. Chapter 8 169--189
\newblock Emerald Group Publishing Limited (2015)

\bibitem{Hambleton1991}
Ronald K. Hambleton and H. Swaminathan and H. Jane Rogers.
\newblock Fundamentals of item response theory.
\newblock Sage Publications: Newbury Park, CA (1991)

\bibitem{inep2010}
\newblock Para entender a nota do Enem.
\newblock Instituto Nacional de Estudos e Pesquisas Educacionais An\'isio Teixeira, Bras\'ilia. http://portal.inep.gov.br (2010). Accessed 12/12/2017

\bibitem{inep2017}
\newblock Reda\c{c}\~ao no Enem 2017: Cartilha do participante.
\newblock Instituto Nacional de Estudos e Pesquisas Educacionais An\'isio Teixeira, Bras\'ilia. http://portal.inep.gov.br (2017)

\bibitem{Soares2006}
Jos\'e Francisco Soares
\newblock Measuring cognitive achievement gaps and inequalities: The case of Brazil
\newblock International Journal of Educational Research 45(3): 176--187 (2006)

\bibitem{Welch1947}
Bernard Lewis Welch.
\newblock The generalization of `Student's' problem when several different population variances are involved.
\newblock Biometrika 34(1-2): 28--35 (1947)

\bibitem{Welch1951}
Bernard Lewis Welch.
\newblock On the comparison of several mean values: An alternative approach.
\newblock Biometrika 38(3-4): 330--336 (1951)

\bibitem{Ruxton2006}
Graeme D. Ruxton.
\newblock The unequal variance t-test is an underused alternative to Student's t-test and the Mann–Whitney U test.
\newblock Behavioral Ecology 17(4): 688--690 (2006)

\bibitem{Jackson2004}
John P. Jackson and Nadine M. Weidman.
\newblock Race, racism, and science: Social impact and interaction.
\newblock ABC-CLIO (2004)

\bibitem{daSilva2000}
Eronides da Silva Lima.
\newblock Mal de fome e n\~ao de ra\c{c}a: G\^enese, constitui\c{c}\~ao e a\c{c}\~ao pol\'itica da educa\c{c}\~ao alimentar: Brasil 1934-1946.
\newblock Editora Fiocruz 285p (2000)

\bibitem{Deary2009}
Ian J. Deary and W. Johnson and L. M. Houlihan.
\newblock Genetic foundations of human intelligence.
\newblock Human Genetics 126(1): 215--232 (2009)

\bibitem{Flynn2007}
James R Flynn.
\newblock What is intelligence? Beyond the Flynn effect.
\newblock Cambridge University Press 274p (2009)   

\bibitem{Dickens2006}
W. T. Dickens and J. R. Flynn.
\newblock Black Americans reduce the racial {IQ} gap: Evidence from standardization samples.
\newblock Psychological Science 17(10): 913--920 (2006)

\bibitem{Weihua2011}
Weihua Niu and Jillian Brass.
\newblock Intelligence in worldwide perspective.
\newblock In R.J. Sternberg and S. B. Kaufmann (Eds.). The Cambridge handbook of intelligence 623--646. Cambridge University Press (2011)

\bibitem{Maluccio2009}
John A Maluccio and John Hoddinott and Jere R. Behrman and Reynaldo Martorell and Agnes R. Quisumbing and Aryeh D. Stein.
\newblock The impact of improving nutrition during early childhood on education among Guatemalan adults.
\newblock The Economic Journal 119(537): 734--763 (2009)

\bibitem{Jeynes2008}
William Jeynes.
\newblock What we should and should not learn from the Japanese and other East Asian education systems.
\newblock Educational Policy 22(6): 900--927 (2008)

\bibitem{Saito1973}
Hiroshi Saito and Takashi Maeyama.
\newblock Assimilação e integração dos Japoneses no Brasil.
\newblock Editora Vozes: Petr\'opolis 558p (1973)

\bibitem{Saviani2005}
Dermeval Saviani.
\newblock As concep\c{c}\~oes pedag\'ogicas na hist\'oria da educa\c{c}\~ao Brasileira.
\newblock Projeto ``20 anos do HISTEDBR'': Publisher Unicamp: Campinas (2005)

\bibitem{daSilva2002}
A\'ilson Bas\'ilio da Silva.
\newblock A hist\'oria da educa\c{c}\~ao no Brasil.
\newblock Monograph. Publisher Universidade Candido Mendes: Rio de Janeiro (2002)

\bibitem{Thomas2017}
Drew Thomas.
\newblock Racial IQ differences among transracial adoptees: Fact or artifact?
\newblock Journal of Intelligence 5:1 (2017)

\bibitem{Pereira2011}
J\'unia Sales Pereira.
\newblock Di\'alogos sobre o exerc\'icio da doc\^encia-recep\c{c}\~ao das leis 10.639/03 e 11.645/08.
\newblock Educa\c{c}\~ao \& Realidade 36(1): 147--172 (2011)

\bibitem{Wood1988}
Charles H. Wood and Jos\'e Alberto De Carvalho.
\newblock The demography of inequality in Brazil.
\newblock Cambridge University Press (1988)

\bibitem{Scarr1976}
Sandra Scarr and Richard A. Weinberg.
\newblock IQ test performance of Black children adopted by White families.
\newblock American Psychologist 31(10): 726--739 (1976)

\bibitem{Ijzendoorn2005}
Marinus H. Van Ijzendoorn and Juffer Femmie and Caroline W. Poelhuis.
\newblock Adoption and cognitive development: A meta-analytic comparison of adopted and nonadopted children's IQ and school performance.
\newblock Psychological Bulletin 131(2): 301--316 (2005)

\bibitem{Dani2005}
Jennifer Dani and Courtney Burrill and Barbara Demmig-Adams.
\newblock The remarkable role of nutrition in learning and behaviour.
\newblock Nutrition \& Food Science 35(4): 258--263 (2005)

\bibitem{Marteleto2013}
Leticia Marteleto and Fernando Andrade
\newblock The Educational Achievement of Brazilian Adolescents -- Cultural Capital and the Interaction between Families and Schools
\newblock Sociology of Education 87(1): 16--35 (2013)

\bibitem{Cloer2001}
Thomas Cloer Jr. and Shana Ross Dalton.
\newblock Gender and grade differences in reading achievement and in self-concept as readers.
\newblock Journal of Reading Education 26(2): 31--36 (2001)

\bibitem{Hyde1990}
Janet S. Hyde and Elizabeth Fennema and Susan J. Lamon.
\newblock Gender differences in mathematics performance: A meta-analysis.
\newblock Psychological Bulletin 107(2): 139-155 (1990)

\bibitem{Guiso2008}
Luigi Guiso and Ferdinando Monte and Paola Sapienza and Luigi Zingales.
\newblock Culture, gender, and math.
\newblock Science 320(5880): 1164--1165 (2008)

\bibitem{Hanushek2010}
Eric A. Hanushek.
\newblock How well do we understand achievement gaps?
\newblock Focus 27(2): 5--12 (2010)

\bibitem{Zhao2016}
D. Zhao.
\newblock Review of the literature: Factors contributing to achievement {GAP}.
\newblock In: Chinese Students' Higher Achievement in Mathematics. Mathematics Education -- An Asian Perspective.
\newblock Springer: Singapore (2016)

\bibitem{Azzolini2012}
Davide Azzolini and Philipp Schnell and John Palmerd.
\newblock Educational achievement gaps between immigrant and native students in two ``New immigration countries'': Italy and Spain in comparison.
\newblock The Annals of the American Academy of Political and Social Science 643(1): 46--77 (2012)

\bibitem{Zhang2011}
Liang Zhang and Kristen A. Lee.
\newblock Decomposing achievement gaps among {OECD} countries.
\newblock Asia Pacific Education Review 12(3): 463--474 (2011)

\end{thebibliography}
\end{document}


\title{Supplementary Information: \\ Assessing student's achievement gap between ethnic groups in Brazil}

\author{Luis E C Rocha}\email{luis.rocha@sociology.su.se}
\affiliation{Department of Sociology, Stockholm University, Stockholm, Sweden}
\author{Luana F Nascimento}\email{ldfnasci@sckcen.be}
\affiliation{Radiation Protection Dosimetry and Calibration group, Belgian Nuclear Research Center, Mol, Belgium}

\date{\today}

\maketitle

\begin{table}[ht]
\centering
\caption{ {\bf Keywords for the inclusion criteria, values and brief description.} }
\label{my-label}
\begin{tabular}{lll}
CODE                      & VALUE         & BRIEF DESCRIPTION                   \\ \hline 
ID\_DEPENDENCIA\_ADM\_ESC & 1 AND 2 AND 3 & school is public management         \\
ID\_LOCALIZACAO\_ESC      & 1             & school is urban                     \\
SIT\_FUNC\_ESC            & 1             & school is active                    \\
IDADE                     & 17            & student age                         \\
NACIONALIDADE             & 1             & student is born in Brazil           \\
ST\_CONCLUSAO             & 2             & student will graduate in 2013       \\
TP\_ESCOLA                & 1             & school is public                    \\
IN\_TP\_ENSINO            & 1             & school is not for special students  \\
TP\_ESTADO\_CIVIL         & 0             & student is single                   \\
IN\_BAIXA\_VISAO          & 0             & student has no low vision           \\
IN\_CEGUEIRA              & 0             & student is not blind                \\
IN\_SURDEZ                & 0             & student is not deaf                 \\
IN\_DEFICIENCIA\_AUDITIVA & 0             & student has no auditory disability  \\
IN\_SURDO\_CEGUEIRA       & 0             & student is not deaf and not blind   \\
IN\_DEFICIENCIA\_FISICA   & 0             & student has no physical disability  \\
IN\_DEFICIENCIA\_MENTAL   & 0             & student has no mental disability    \\
IN\_DEFICIT\_ATENCAO      & 0             & student has no attention deficit    \\
IN\_DISLEXIA              & 0             & student has not dyslexia            \\
IN\_GESTANTE              & 0             & student is not pregnant             \\
IN\_LACTANTE              & 0             & student is not lactating            \\                                 
IN\_IDOSO                   & 0           & elderly student, 60 or above                                         \\
IN\_AUTISMO                 & 0           & student is not autistic                                              \\
IN\_BRAILLE                 & 0           & student does not need exam in Braille                                \\
IN\_AMPLIADA\_24            & 0           & student does not need exam with larger fonts \\
IN\_AMPLIADA\_18            & 0           & student does not need exam with larger fonts \\
IN\_LEDOR                   & 0           & student does not need an auxiliary person to read and write the exam \\
IN\_ACESSO                  & 0           & student does not need a class with special access                    \\
IN\_TRANSCRICAO             & 0           & student does not need an auxiliary person to transcribe the text     \\
IN\_LIBRAS                  & 0           & student does not need an auxiliary person for sign language          \\
IN\_LEITURA\_LABIAL         & 0           & student does not need an auxiliary person for lip reading            \\
IN\_MESA\_CADEIRA\_RODAS    & 0           & student does not need a desk for wheelchair                          \\
IN\_MESA\_CADEIRA\_SEPARADA & 0           & student does not need separated desk and chair                       \\
IN\_APOIO\_PERNA            & 0           & student does not need an auxiliary device to rest his/her legs       \\
IN\_GUIA\_INTERPRETE        & 0           & student does not need an auxiliary person for translation            \\
IN\_PRESENCA\_CN            & 1           & student participated in the natural sciences examination             \\
IN\_PRESENCA\_CH            & 1           & student participated in the human sciences examination               \\
IN\_PRESENCA\_LC            & 1           & student participated in the languages examination                    \\
IN\_PRESENCA\_MT            & 1           & student participated in the mathematics examination                  \\
Q004                        & 1 AND 2 AND 3 & student lives in a household with 6 or less people                   \\
                        & 4 AND 5 AND 6 & \\

Q006                        & B           & student lives in an urban area   \\ \hline
\end{tabular}
\end{table}

\begin{table}[ht]
\rowcolors{1}{}{verylightgray}
\setlength\tabcolsep{2.0pt}
\centering
\caption{ {\bf The mean performance scores and $p$-values for household income levels in mathematics. Very low income corresponds to G1 in the main text, and very high income corresponds to G6. Cases of $p<.01$ are highlighted.}}
\begin{tabular}{cccccccccccc}
\rowcolor{verylightgray}
          &                    &         \multicolumn{5}{c}{Very low income} & \multicolumn{5}{c}{Very high income} \\
State &                    & All    & W-B  & W-P & W-Y & W-I  & All & W-B & W-P & W-Y & W-I \\ \hline

AC & $\Delta (\%)$ &         & 0.9   & -2.0  & -3.3  & -   &   & - & - & - & - \\
      & p-value          & .40   & .73 & .29 & .46 & -       & - & - & -        & - & -  \\

AL & $\Delta (\%)$ &         & -0.2 & 1.3    & -1.9 & 4.5   &   & - & - & - & - \\
     & p-value          & .47 & .91 & .29 & .43 & .43 & - & - & - & - & - \\

AP & $\Delta (\%)$ &         & 0.1 & -0.3 & -7.0 & - & & - & - & - & - \\
     & p-value           & .43 & .96 & .86 & .13 & - & - & - & - & - & - \\

AM & $\Delta (\%)$ &         & -1.0 & 0.3 & -2.2 & -2.0 & & - & 0.6 & - & - \\
     & p-value           & .67 & .57 & .79 & .38 & .50 & - & - & .89 & - & - \\

BA & $\Delta (\%)$ &         &1.4 & \cellcolor{lightgray} 1.7 & 1.2 & 2.6 &  & - & 2.6 & - & - \\
     & p-value           & .09 & .04 & \cellcolor{lightgray} $<.01$ & .35 & .17 & - & - & .71 & - & - \\

CE & $\Delta (\%)$ & & 0.6 & \cellcolor{lightgray} 2.7 & -0.9 & 1.2 & & - & 12.0 & - & - \\
     & p-value           & \cellcolor{lightgray} $<.01$ & .40 &  \cellcolor{lightgray} $<.01$ & .53 & .64 & - & - & .25 & - & - \\

DF & $\Delta (\%)$ & & 3.7 & 3.3 & -0.1 & 11.0 & & 10.3 & 5.4 & - & - \\
     & p-value           & .11 & .07 & .03 & .98 & - & \cellcolor{lightgray} $<.01$ & .01 & .01 & - & - \\

ES & $\Delta (\%)$ & & \cellcolor{lightgray} 4.1 & 1.2 & 0.1 & -4.2 & & - & 2.0 & - & - \\
     & p-value           & .08 & \cellcolor{lightgray} $<.01$ & .23 & .97 & .34 & - & - & .65 & - & - \\

GO & $\Delta (\%)$ & & 1.0 & 1.5 & 0.9 & 2.3 & & -2.3 & -0.2 & - & - \\
     & p-value            & .35 & .33 & .04 & .60 & .48 & .94 & .73 & .94 & - & - \\

MA & $\Delta (\%)$ & & 0.4 & 1.3 & -1.3 & 6.9 & & - & - & - & - \\
     & p-value           & .05 & .59 & .05 & .48 & .04 & - & - & - & - & - \\

MT & $\Delta (\%)$ & & 1.1 & 1.8 & 4.8 & - & & - & 0.2 & - & - \\
     & p-value            & .31 & .45 & .09 & .25 & - & - & - & .82 & - & - \\

MS & $\Delta (\%)$ & & -0.6 & 1.9 & -1.4 & - & & - & -1.1 & - & - \\
     & p-value           & .21 & .77 & .06 & .72 & - & - & - & .84 & - & - \\

MG & $\Delta (\%)$ & & \cellcolor{lightgray} 2.9 & \cellcolor{lightgray} 2.6 & \cellcolor{lightgray} 3.4 & 3.3 & & 5.2 & 2.0 & 13.7 & - \\
     & p-value            & \cellcolor{lightgray} $<.01$ & \cellcolor{lightgray} $<.01$ & \cellcolor{lightgray} $<.01$ & \cellcolor{lightgray} $<.01$ & .06 & .05 & .05 & .21 & .05 & - \\

PR & $\Delta (\%)$ & & 4.1 & 1.1 & 1.3 & - & & 1.0 & 0 & -11.0 & - \\
     & p-value            & .01 & <.01 & .09 & .60 & - & .46 & .89 & .99 & .13 & - \\

PB & $\Delta (\%)$ & & 2.9 & 1.6 & 1.9 & 3.6 & & - & - & - & - \\
     & p-value           & .10 & .01 & .03 & .30 & .29 & - & - & - & - & - \\

PA & $\Delta (\%)$ & & -1.7 & 0.1 & 2.0 & 2.4 & & - & 2.9 & - & - \\
     & p-value           & .28 & .13 & .93 & .37 & .58 & - & - & .58 & - & - \\

PE & $\Delta (\%)$ & & 1.2 & 1.2 & 1.8 & 3.7 & & - & 4.6 & - & - \\
     & p-value           & .07 & .09 & .01 & .16 & .09 & - & - & .36 & - & - \\

PI  & $\Delta (\%)$& & 1.1 & 1.6 & -0.6 & 2.8 & & - & - & - & - \\
     & p-value          & .48 & .42 & .10 & .79 & .63 & - & - & - & - & - \\

RJ & $\Delta (\%)$ & & \cellcolor{lightgray} 4.5 & \cellcolor{lightgray} 3.2 & 1.4 & 5.2 & & 11.2 & \cellcolor{lightgray} 7.0 & - & - \\
& p-value          & \cellcolor{lightgray} $<.01$ &\cellcolor{lightgray}  $<.01$ & \cellcolor{lightgray} $<.01$ & .51 & .24 & \cellcolor{lightgray} $<.01$ & .01 & \cellcolor{lightgray} $<.01$ & - & - \\

RN & $\Delta (\%)$ & & 0 & 0.9 & 2.5 & 0 & & - & - & - & - \\
& p-value          & .71 & .99 & .28 & .28 & .99 & - & - & - & - & - \\

RO & $\Delta (\%)$ & & -4.5 & -2.0 & -1.6 & - & & - & 5.3 & - & - \\
& p-value          & .23 & .04 & .19 & .60 & - & - & - & .31 & - & - \\

RR & $\Delta (\%)$ & & 0.4 & 0.5 & - & - & & - & - & - & - \\
& p-value          & .98 & .93 & .84 & - & - & - & - & - & - & - \\

RS & $\Delta (\%)$ & & \cellcolor{lightgray} 3.6 & 2.1 & 0.6 & - & & 6.2 & 2.4 & - & - \\
& p-value          & \cellcolor{lightgray} $<.01$ & \cellcolor{lightgray} $<.01$ & .01 & .85 & - & .44 & .26 & .51 & - & - \\

SC & $\Delta (\%)$ & & 1.4 & 2.7 & -3.4 & - & & - & 3.4 & - & - \\
& p-value          & .10 & .47 & .02 & .39 & - & - & - & .38 & - & - \\

SE & $\Delta (\%)$ & & -1.0 & -0.8 & 0.4 & 3.0 & & - & -1- & - & - \\
& p-value          & .83 & .58 & .58 & .89 & .45 & - & - & - & - & - \\

SP & $\Delta (\%)$ & & \cellcolor{lightgray} 2.0 & \cellcolor{lightgray} 2.1 & 1.0 & 1.8 & & \cellcolor{lightgray} 6.5 & \cellcolor{lightgray} 3.9 & \cellcolor{lightgray} -14.5 & - \\
& p-value          & \cellcolor{lightgray} $<.01$ & \cellcolor{lightgray} $<.01$ & \cellcolor{lightgray} $<.01$ & .43 & .34 & \cellcolor{lightgray} $<.01$ & \cellcolor{lightgray} $<.01$ & \cellcolor{lightgray} $<.01$ & \cellcolor{lightgray} $<.01$ & - \\

TO & $\Delta (\%)$ & &  2.6 & 0.3 & -3.4 & - & & -7.2 & -7.4 & - & - \\
& p-value          & .10 & .10 & .81 & .21 & - & .44 & .33 & .21 & - & - \\ \hline
\end{tabular}
\\ W: White; B: Black; P: Pardo; Y: Yellow; I: Indigenous; All: All ethnic groups compared simultaneously.
\label{tab:01}
\end{table}

\begin{table}[ht]
\rowcolors{1}{}{verylightgray}
\setlength\tabcolsep{2.0pt}
\centering
\caption{
{\bf The mean performance scores and $p$-values for household income levels in writing. Very low income corresponds to G1 in the main text, and very high income corresponds to G6. Cases of $p<.01$ are highlighted.}}
\begin{tabular}{cccccccccccc}
\rowcolor{verylightgray}
          &                    &         \multicolumn{5}{c}{Very low income} & \multicolumn{5}{c}{Very high income} \\
State &                    & All    & W-B  & W-P & W-Y & W-I  & All & W-B & W-P & W-Y & W-I \\ \hline

AC & $\Delta (\%)$ & & -4.4 & -0.2 & -11.1 & - & & - & - & - & - \\
& p-value         & .10 & .39 & .96 & .04 & - & - & - & - & - & - \\

AL & $\Delta (\%)$ & & 2.6 & 1.9 & -0.6 & 2.5 & & - & - & - & - \\
& p-value         & .93 & .52 & .49 & .91 & .84 & - & - & - & - & - \\

AP & $\Delta (\%)$ & & \cellcolor{lightgray} 13.5 & 8.8 & -8.6 & - & & - & - & - & - \\
& p-value         & \cellcolor{lightgray}$<.01$ & \cellcolor{lightgray}$<.01$ & .02 & .28 & - & - & - & - & - & - \\

AM & $\Delta (\%)$ & & 5.0 & \cellcolor{lightgray} 6.2 & -7.3 & 4.0 & & - & 1.0 & - & - \\
& p-value         & \cellcolor{lightgray} $<.01$ & .22 & \cellcolor{lightgray} $<.01$ & .15 & .54 & - & - & .83 & - & - \\

BA & $\Delta (\%)$ & & 3.6 & 3.0 & -0.6 & \cellcolor{lightgray} 16.8 & & - & -2.6 & - & - \\
& p-value         & \cellcolor{lightgray} $<.01$ & .01 & .02 & .83 & \cellcolor{lightgray} $<.01$ & - & - & .70 & - & - \\

CE & $\Delta (\%)$ & & 1.6 & \cellcolor{lightgray} 5.6 & 0.6 & 4.7 & & - & 7.3 & - & - \\
& p-value         & \cellcolor{lightgray} $<.01$ & .27 & \cellcolor{lightgray} $<.01$ & .81 & .36 & - & - & .56 & - & - \\

DF & $\Delta (\%)$ & & 0.1 & 4.1 & 1.9 & - & & 10.0 & 5.6 & - & - \\
& p-value         & .35 & .98 & .11 & .73 & - & .01 & .01 & .02 & - & - \\

ES & $\Delta (\%)$ & & 7.5 & 3.9 & 8.9 & 4.3 & &  - & 4.2 & - & - \\
& p-value         & .09 & .01 & .03 & .24 & .58 & - & - & .46 & - & - \\

GO & $\Delta (\%)$ & & 1.6 & 3.2 & -2.4 & 1.1 & & -1.7 & 4.2 & - & - \\
& p-value         & .13 & .47 & .04 & .41 & .88 & .59 & .83 & .34 & - & - \\

MA & $\Delta (\%)$ &          & 1.3   & \cellcolor{lightgray} 4.2         & -2.8  & -5.0  &   & - & - & - & - \\
& p-value          & .03 & .52 & \cellcolor{lightgray} $<.01$ & .51 & .60 & - & -       & -     & -      & - \\

MT & $\Delta (\%)$ & & \cellcolor{lightgray} 9.0 & 3.7 & \cellcolor{lightgray} 23.1 & - & & - & - & - & - \\
& p-value         & \cellcolor{lightgray} $<.01$ & \cellcolor{lightgray} $<.01$ & .09 & \cellcolor{lightgray} $<.01$ & - & - & - & - & - & - \\

MS & $\Delta (\%)$ & & 1.3 & 3.0 & 23.7 & - & & - & 0.2 & - & - \\
& p-value         & .11 & .72 & .12 & .05 & - & - & - & .98 & - & - \\

MG & $\Delta (\%)$ & & \cellcolor{lightgray} 4.7 & \cellcolor{lightgray} 4.9 & 4.6 & 6.0 & & 2.9 & 2.4 & 11.2 & - \\
& p-value         & \cellcolor{lightgray} $<.01$ & \cellcolor{lightgray} $<.01$ & \cellcolor{lightgray} $<.01$ & .04 & .09 & .20 & .45 & .23 & .07 & - \\

PR & $\Delta (\%)$ & & \cellcolor{lightgray} 9.3 & 2.9 & -6.1 & - & & 7.8 & 7.3 & -9.8 & - \\
& p-value         & \cellcolor{lightgray}$<.01$ & \cellcolor{lightgray} $<.01$ & .03 & .20 & - & .11 & .39 & .05 & .19 & - \\

PB & $\Delta (\%)$ & & 2.4 & 1.6 & 4.3 & 7.8 & & - & - & - & - \\
& p-value         & .56 & .31 & .30 & .31 & .25 & - & - & - & - & - \\

PA & $\Delta (\%)$ & & -0.8 & 2.2 & -1.8 & 4.9 & & - & -0.3 & - & - \\
& p-value         & .28 & .70 & .17 & .69 & .65 & - & - & .98 & - & - \\

PE & $\Delta (\%)$ & & \cellcolor{lightgray} 6.3 & \cellcolor{lightgray} 5.2 & \cellcolor{lightgray} 8.0 & 5.5 & & - & 9.2 & - & - \\
& p-value         & \cellcolor{lightgray} $<.01$ & \cellcolor{lightgray} $<.01$ & \cellcolor{lightgray} $<.01$ & \cellcolor{lightgray} $<.01$ & .17 & - & - & .11 & - & - \\

PI & $\Delta (\%)$ & & 4.9 & 2.8 & 1.6 & 8.1 & & - & - & - & - \\
& p-value         & .49 & .08 & .18 & .73 & .44 & - & - & - & - & - \\

RJ & $\Delta (\%)$ & & \cellcolor{lightgray} 5.5 & \cellcolor{lightgray} 6.1 & -2.5 & -0.7 & & 5.2 & 4.0 & - & - \\
& p-value         & \cellcolor{lightgray} $<.01$ & \cellcolor{lightgray} $<.01$ & \cellcolor{lightgray} $<.01$ & .44 & .92 & .14 & .18 & .09 & - & - \\

RN & $\Delta (\%)$ & & 4.1 & 3.0 & 1.5 & -3.8 & & - & - & - & - \\
& p-value         & .36 & .12 & .08 & .77 & .63 & - & - & - & - & - \\

RO & $\Delta (\%)$ & & -3.3 & 0.3 & 1.2 & - & &  - & -6.7 & - & - \\
& p-value         & .89 & .51 & .93 & .90 & - & - & - & .54 & - & - \\

RR & $\Delta (\%)$ & & 10.0 & 12.5 & - & - & & - & - & - & - \\
& p-value         & .11 & .22 & .04 & - & - & - & - & - & - & - \\

RS & $\Delta (\%)$ & & 2.4 & 2.7 & 4.8 & - & & 15.9 & -3.6 & - & - \\
& p-value         & .23 & .24 & .08 & .44 & - & .12 & .08 & .31 & - & - \\

SC & $\Delta (\%)$ & & -1.2 & 0 & -0.2 & - & & - & -3.4 & - & - \\
& p-value         & .99 & .71 & .99 & .94 & - & .99 & - & .48 & - & - \\

SE & $\Delta (\%)$ & & -3.3 & -5.7 & -9.1 & 9.0 & & - & - & - & - \\
& p-value         & .20 & .39 & .06 & .15 & .43 & - & - & - & - & - \\

SP & $\Delta (\%)$ & & \cellcolor{lightgray} 3.4 & \cellcolor{lightgray} 3.4 & 2.2 & 1.3 & & \cellcolor{lightgray} 9.0 & 2.6 & \cellcolor{lightgray} -7.4 & - \\
& p-value         & \cellcolor{lightgray} $<.01$ & \cellcolor{lightgray} $<.01$ & \cellcolor{lightgray} $<.01$ & .42 & .72 & \cellcolor{lightgray} $<.01$ & \cellcolor{lightgray} $<.01$ & .04 &\cellcolor{lightgray} $<.01$ & - \\

TO & $\Delta (\%)$ & & 5.4 & 1.3 & 0.9 & - & & 8.4 & 6.7 & - & - \\
& p-value         & .58 & .21 & .69 & .89 & - & .40 & .19 & .28 & - & - \\ \hline
\end{tabular}
\\ W: White; B: Black; P: Pardo; Y: Yellow; I: Indigenous; All: All ethnic groups compared simultaneously.
\label{tab:02}
\end{table}

\begin{table}[ht]
\rowcolors{1}{}{verylightgray}
\setlength\tabcolsep{2.0pt}
\centering
\caption{
{\bf The mean performance scores and $p$-values for parental education levels in mathematics. Very low education corresponds to G2 in the main text, and very high education corresponds to G5. Cases of $p<.01$ are highlighted.}}
\begin{tabular}{cccccccccccc}
\rowcolor{verylightgray}
          &                    &         \multicolumn{5}{c}{Very low education} & \multicolumn{5}{c}{Very high education} \\
State &                    & All    & W-B  & W-P & W-Y & W-I  & All & W-B & W-P & W-Y & W-I \\ \hline

AC & $\Delta (\%)$ &         & 0.6  & 5.4  & -  & -   &   & 2.6 & -3.0 & - & - \\
      & p-value          & .2       & .90   & .15  & - & -    & .11 & .44 & .17 & - & -  \\

AL & $\Delta (\%)$ &         & \cellcolor{lightgray} 8.2 & 4.0 & 2.5 & -   &   & -3.4 & 3.7 & - & - \\
     & p-value          & .08 & \cellcolor{lightgray} $<.01$ & .07 & .61 & - & .27 & .56 & .23 & - & - \\

AP & $\Delta (\%)$ &         & - & -1.1 & - & - & & -0.5 & -2.0 & - & - \\
     & p-value           & - & - & .88 & - & - & .65 & .88 & 0.39 & - & - \\

AM & $\Delta (\%)$ &         & -0.1 & 5.6 & - & - & & -1.4 & 3.2 & -4.7 & 8.1 \\
     & p-value           & .09 & .98 & .08 & - & - & .04 & .62 & .02 & .34 & .10 \\

BA & $\Delta (\%)$ &         & 3.7 & 2.4 & 3.2 & -1.6 &  & 0.1 & 0 & 3.9 & 5.0 \\
     & p-value           & .14 & .01 & .06 & .25 & .73 & .54 & .54 & .98 & .19 & .35 \\

CE & $\Delta (\%)$ & & -0.5 & \cellcolor{lightgray} 3.0 & -2.5 & \cellcolor{lightgray} 14.0 & & -0.1 & 0.9 & -1.2 & 4.9 \\
     & p-value           & \cellcolor{lightgray} $<.01$ & .74 & \cellcolor{lightgray} $<0.01$ & .32 & \cellcolor{lightgray} $<.01$ & .75 & .97 & .49 & .74 & .27 \\

DF & $\Delta (\%)$ & & -1.0 & 3.8 & - & - & & \cellcolor{lightgray} 6.2 & \cellcolor{lightgray} 4.4 & 2.8 & - \\
     & p-value           & .2 & .80 & .17 & - & - & \cellcolor{lightgray} $<.01$ & \cellcolor{lightgray} $<.01$ & \cellcolor{lightgray} $<.01$ & 0.44 & - \\

ES & $\Delta (\%)$ & & \cellcolor{lightgray} 6.2 & \cellcolor{lightgray} 4.5 & - & - & & 4.8 & 2.6 & 4.9 & - \\
     & p-value           & \cellcolor{lightgray} $<.01$ & \cellcolor{lightgray} $<.01$ & \cellcolor{lightgray} $<.01$ & - & - & .06 & .02 & .04 & .34 & - \\

GO & $\Delta (\%)$ & & 4.1 & 1.4 & 1.8 & - & & 3.5 & 1.8 & 2.8 & - \\
     & p-value            & .18 & .03 & .29 & .43 & - & .07 & .02 & .05 & .20 & - \\

MA & $\Delta (\%)$ & & 2.8 & 0.8 & 2.9 & - & & -1.7 & 0.9 & 2.6 & - \\
     & p-value           & .46 & .14 & .57 & .43 & - & .48 & .39 & .46 & .57 & - \\

MT & $\Delta (\%)$ & & 0 & 3.8 & - & - & & \cellcolor{lightgray} 4.2 & 2.3 & 2.4 & - \\
     & p-value            & .08 & .99 & .04 & - & - & .04 & \cellcolor{lightgray} $<.01$ & .02 & .68 & - \\

MS & $\Delta (\%)$ & & -3.2 & 4.3 & - & - & & 2.7 & 2.7 & -3.2 & - \\
     & p-value           & .04 & .36 & .04 & - & - & .14 & .36 & .03 & .55 & - \\

MG & $\Delta (\%)$ & & \cellcolor{lightgray} 6.3 & \cellcolor{lightgray} 3.8 & \cellcolor{lightgray} 4.8 & 4.9 & & \cellcolor{lightgray} 4.1 & \cellcolor{lightgray} 3.3 & \cellcolor{lightgray} 5.2 & 9.5 \\
     & p-value            & \cellcolor{lightgray} $<.01$ & \cellcolor{lightgray} $<.01$ & \cellcolor{lightgray} $<.01$ & \cellcolor{lightgray} $<.01$ & .13 & \cellcolor{lightgray} $<.01$ & \cellcolor{lightgray} $<.01$ & \cellcolor{lightgray} $<.01$ & \cellcolor{lightgray} $<.01$ & .02 \\

PR & $\Delta (\%)$ & & 3.9 & 1.7 & -5.6 & - & & \cellcolor{lightgray} 5.7 & \cellcolor{lightgray} 3.9 & \cellcolor{lightgray} -8.4 & - \\
     & p-value            & .01 & .01 & .02 & .29 & - & \cellcolor{lightgray} $<.01$ & \cellcolor{lightgray} $<.01$ & \cellcolor{lightgray} $<.01$ & \cellcolor{lightgray} $<.01$ & - \\

PB & $\Delta (\%)$ & & -0.8 & -0.3 & -1.0 & 7.9 & & 4.9 & \cellcolor{lightgray} 5.3 & 11.0 & - \\
     & p-value           & .14 & .74 & .82 & .78 & .03 & \cellcolor{lightgray} $.<01$ & 0.10 & \cellcolor{lightgray} $<.01$ & .01 & - \\

PA & $\Delta (\%)$ & & 1.7 & 1.6 & - & - & & -1.4 & 0.6 & 1.2 & - \\
     & p-value           & .75 & .54 & .45 & - & - & .76 & .53 & .68 & .79 & - \\

PE & $\Delta (\%)$ & & 2.7 & 0.6 & 0.3 & -2.4 & & \cellcolor{lightgray} 6.0 & 1.9 & 4.9 & 9.2 \\
     & p-value           & .56 & .10 & .57 & .90 & .70 & \cellcolor{lightgray} $<.01$ & \cellcolor{lightgray} $<.01$ & .09 & .16 & .02 \\

PI  & $\Delta (\%)$& & 3.7 & 4.5 & 6.1 & - & & 1.8 & -1.3 & -3.5 & - \\
     & p-value          & .11 & .20 & .02 & .12 & - & .50 & .49 & .51 & .51 & - \\

RJ & $\Delta (\%)$ & & 2.0 & -0.9 & 4.2 & - & & \cellcolor{lightgray} 10.0 & 5.9 & \cellcolor{lightgray} 17.2 & - \\
& p-value          & .30 & .29 & .56 & .25 & - & \cellcolor{lightgray} $<.01$ & \cellcolor{lightgray} $<.01$ & $<.01$ & \cellcolor{lightgray} $<.01$ & - \\

RN & $\Delta (\%)$ & & 1.2 & 2.8 & 7.9 & - & & 2.0 & 2.1 & 0.2 & - \\
& p-value          & .18 & .68 & .10 & .07 & - & .81 & .64 & .33 & .98 & - \\

RS & $\Delta (\%)$ & & \cellcolor{lightgray} 7.8 & 3.1 & 7.9 & - & & \cellcolor{lightgray} 8.5 & \cellcolor{lightgray} 3.8 & 5.5 & - \\
& p-value          & \cellcolor{lightgray} $<.01$ & \cellcolor{lightgray} $<.01$ & .01 & .09 & - & \cellcolor{lightgray} $<.01$ & \cellcolor{lightgray} $<.01$ & \cellcolor{lightgray} $<.01$ & .21 & - \\

RO & $\Delta (\%)$ & & -0.4 & -0.1 & - & - & & -0.8 & 1.6 & 4.5 & - \\
& p-value          & - & .92 & .97 & - & - & .48 & .80 & .25 & .27 & - \\

RR & $\Delta (\%)$ & & - & - & - & - & & -1.9 & 1.8 & - & - \\
& p-value          & - & - & - & - & - & .64 & .68 & .56 & - & - \\

SC & $\Delta (\%)$ & & 3.3 & 1.8 & -1.4 & - & & 4.8 & \cellcolor{lightgray} 5.6 & 6.8 & - \\
& p-value          & .24 & .20 & .11 & .62 & - & \cellcolor{lightgray} $<.01$ & .07 & \cellcolor{lightgray} $<.01$ & .06 & - \\

SE & $\Delta (\%)$ & & -2.8 & -2.2 & 0.6 & - & & 1.7 & 0.9 & - & - \\
& p-value          & .81 & .45 & .47 & .90 & - & .93 & .71 & .82 & - & - \\

SP & $\Delta (\%)$ & & \cellcolor{lightgray} 2.9 & \cellcolor{lightgray} 2.6 & 2.8 & 4.8 & & \cellcolor{lightgray} 6.2 & \cellcolor{lightgray} 5.5 & \cellcolor{lightgray} -9.1 & 7.3 \\
& p-value          & \cellcolor{lightgray} $<.01$ & \cellcolor{lightgray} $<.01$ & \cellcolor{lightgray} $<.01$ & .11 & .03 & \cellcolor{lightgray} $<.01$ & \cellcolor{lightgray} $<.01$ & \cellcolor{lightgray} $<.01$ & \cellcolor{lightgray} $<.01$ & .01 \\

TO & $\Delta (\%)$ & &  4.7 & 0.4 & - & - & & 3.0 & 5.0 & 2.2 & - \\
& p-value          & .24 & .18 & .89 & - & - & .02 & .16 & $<.01$ & .56 & - \\ \hline

\end{tabular}
\\ W: White; B: Black; P: Pardo; Y: Yellow; I: Indigenous; All: All ethnic categories taken together.
\label{tab:03}
\end{table}

\begin{table}[ht]
\rowcolors{1}{}{verylightgray}
\setlength\tabcolsep{2.0pt}
\centering
\caption{
{\bf The mean performance scores and $p$-values for parental education levels in writing. Very low education corresponds to G2 in the main text, and very high education corresponds to G5. Cases of $p<.01$ are highlighted.}}
\begin{tabular}{cccccccccccc}
\rowcolor{verylightgray}
          &                    &         \multicolumn{5}{c}{Very low education} & \multicolumn{5}{c}{Very high education} \\
State &                    & All    & W-B  & W-P & W-Y & W-I  & All & W-B & W-P & W-Y & W-I \\ \hline
AC & $\Delta (\%)$ &         & -1.7  & 4.4  & -  & -   &   & 1.1 & -0.1 & - & - \\
      & p-value          & .57 & .82  & .45 & - & -    & 0.99 & 0.90 & 0.97 & - & -  \\

AL & $\Delta (\%)$ &         & 11.5 & 4.8 & 2.1 & -   &   & -3.4 & 3.7 & - & - \\
     & p-value          & 0.41 & 0.09 & 0.35 & 0.85 & - & 0.08 & 0.74 & 0.05 & - & - \\

AP & $\Delta (\%)$ &         & - & -11.7 & - & - & & 3.7 & 0.2 & - & - \\
     & p-value           & - & - & 0.43 & - & - & 0.72 & 0.50 & 0.96 & - & - \\

AM & $\Delta (\%)$ &         & -0.5 & 1.5 & - & - & & -3.9 & 1.5 & -1.2 & 5.4 \\
     & p-value           & .96 & .97 & .85 & - & - & .86 & .43 & .41 & .85 & .47 \\

BA & $\Delta (\%)$ &         & 3.2 & -0.7 & -9.9 & 15.6 &  & -3.3 & -2.1 & 1.8 & 20.3 \\
     & p-value           & .12 & .36 & .82 & .09 & .23 & .27 & .12 & .25 & .67 & .22 \\

CE & $\Delta (\%)$ & & 1.4 & \cellcolor{lightgray} 6.0 & -4.2 & -5.4 & & -3.3 & -1.3 & -7.5 & -1.9 \\
     & p-value           & \cellcolor{lightgray} $<.01$ & .69 & \cellcolor{lightgray} $<0.01$ & .40 & .46 & .31 & .25 & .46 & .04 & .74 \\

DF & $\Delta (\%)$ & & -2.1 & 4.2 & - & - & & 0.7 & 3.3 & -0.9 & - \\
     & p-value           & .32 & .66 & .31 & - & - & .17 & .77 & .03 & .84 & - \\

ES & $\Delta (\%)$ & & 9.0 & 2.6 & - & - & & 3.5 & 2.5 & 11.7 & - \\
     & p-value           & .09 & .03 & .31 & - & - & .29 & .19 & .16 & .22 & - \\

GO & $\Delta (\%)$ & & 5.0 & 6.0 & 7.6 & - & & 4.0 & 2.7 & -2.2 & - \\
     & p-value            & .21 & .20 & .04 & .14 & - & .11 & .09 & .06 & .52 & - \\

MA & $\Delta (\%)$ & & -3.80 & 5.2 & -0.7 & - & & -3.1 & -0.4 & -5.4 & - \\
     & p-value           & .13 & .53 & .22 & .95 & - & .60 & .33 & .85 & .33 & - \\

MT & $\Delta (\%)$ & & 5.9 & 3.9 & - & - & & -1.8 & 2.5 & 5.5 & - \\
     & p-value            & .45 & .33 & .27 & - & - & .27 & .51 & .14 & .44 & - \\

MS & $\Delta (\%)$ & & -3.9 & -0.9 & - & - & & -2.9 & 4.5 & 1.7 & - \\
     & p-value           & .78 & .48 & .80 & - & - & .06 & .42 & .01 & .83 & - \\

MG & $\Delta (\%)$ & & \cellcolor{lightgray} 8.1 & \cellcolor{lightgray} 5.5 & 2.2 & 13.1 & & \cellcolor{lightgray} 4.2 & \cellcolor{lightgray} 3.5 & 5.7 & \cellcolor{lightgray} 15.3 \\
     & p-value            & \cellcolor{lightgray} $<.01$ & \cellcolor{lightgray} $<.01$ &\cellcolor{lightgray} $<.01$ & .44 & .02 & \cellcolor{lightgray} $<.01$ & \cellcolor{lightgray} $<.01$ & \cellcolor{lightgray} $<.01$ & .06 & \cellcolor{lightgray} $<.01$ \\

PR & $\Delta (\%)$ & & 7.8 & 2.3 & \cellcolor{lightgray} -18.3 & - & & 1.2 & 2.5 & \cellcolor{lightgray} -9.0 & - \\
     & p-value            & $<.01$ & .03 & .11 & \cellcolor{lightgray} $<.01$ & - & \cellcolor{lightgray} $<.01$ & .58 & .02 & \cellcolor{lightgray} $<.01$ & - \\

PB & $\Delta (\%)$ & & 1.5 & 4.3 & 3.4 & 47.3 & & 8.7 & 1.2 & -3.0 & - \\
     & p-value           & .08 & .72 & .18 & .67 & .02 & .24 & 0.05 & .66 & .75 & - \\

PA & $\Delta (\%)$ & & -2.8 & -0.3 & - & - & & \cellcolor{lightgray} 10.8 & \cellcolor{lightgray} 7.1 & 19.0 & - \\
     & p-value           & .84 & .61 & .95 & - & - & \cellcolor{lightgray} $<.01$ & \cellcolor{lightgray} $<.01$ & \cellcolor{lightgray} $<.01$ & .02 & - \\

PE & $\Delta (\%)$ & & 7.1 & \cellcolor{lightgray} 6.2 & 2.7 & 5.9 & & 4.9 & 2.6 & 6.3 & 14.6 \\
     & p-value           & .07 & .06 & \cellcolor{lightgray} $<.01$ & .55 & .62 & .06 & .06 & .11 & .19 & .03 \\

PI  & $\Delta (\%)$& & 3.7 & 4.0 & 9.2 & - & & 6.7 & -0.2 & -4.9 & - \\
     & p-value          & .81 & .52 & .41 & .43 & - & .39 & .19 & .95 & .54 & - \\

RJ & $\Delta (\%)$ & & 2.3 & -0.5 & 0.5 & - & & \cellcolor{lightgray} 5.3 & \cellcolor{lightgray} 5.4 & 12.6 & - \\
     & p-value          & .83 & .45 & .81 & .94 & - & \cellcolor{lightgray} $<.01$ & \cellcolor{lightgray} $<.01$ & \cellcolor{lightgray} $<.01$ & .04 & - \\

RN & $\Delta (\%)$ & & 8.9 & 6.4 & -6.3 & - & & 0.2 & 2.6 & -3.2 & - \\
     & p-value          & .20 & .18 & .10 & .48 & - & .81 & .98 & .42 & .69 & - \\

RS & $\Delta (\%)$ & & \cellcolor{lightgray} 13.2 & 2.6 & 15.6 & - & & \cellcolor{lightgray} 9.1 & 0.7 & 8.3 & - \\
& p-value          & .02 & \cellcolor{lightgray} $<.01$ & .21 & .11 & - & \cellcolor{lightgray} $<.01$ & \cellcolor{lightgray} $<.01$ & .64 & .24 & - \\

RO & $\Delta (\%)$ & & -10.5 & -1.5 & - & - & & 4.9 & 1.9 & -0.7 & - \\
& p-value          & .29 & .12 & .74 & - & - & .64 & .26 & .39 & .91 & - \\

RR & $\Delta (\%)$ & & - & - & - & - & & -4.4 & 11.6 & - & - \\
& p-value          & - & - & - & - & - & .01 & .53 & .01 & - & - \\

SC & $\Delta (\%)$ & & 2.4 & 1.3 & 0.2 & - & & 5.1 & \cellcolor{lightgray} 6.1 & 0.9 & - \\
& p-value          & .89 & .60 & .52 & .97 & - & \cellcolor{lightgray} $<.01$ & .15 & \cellcolor{lightgray} $<.01$ & .84 & - \\

SE & $\Delta (\%)$ & & 1.4 & -5.4 & -8.6 & - & & 0.6 & -5.5 & - & - \\
& p-value          & .59 & .88 & .43 & .37 & - & .39 & .93 & .32 & - & - \\

SP & $\Delta (\%)$ & & \cellcolor{lightgray} 5.6 & \cellcolor{lightgray} 3.0 & 8.4 & 11.9 & & \cellcolor{lightgray} 5.5 & \cellcolor{lightgray} 5.1 & \cellcolor{lightgray} -3.8 & 5.6 \\
& p-value          & \cellcolor{lightgray} $<.01$ & \cellcolor{lightgray} $<.01$ & \cellcolor{lightgray} $<.01$ & .02 & .03 & \cellcolor{lightgray} $<.01$ & \cellcolor{lightgray} $<.01$ & \cellcolor{lightgray} $<.01$ & \cellcolor{lightgray} $<.01$ & .08 \\

TO & $\Delta (\%)$ & &  11.8 & 13.8 & - & - & & \cellcolor{lightgray} 12.1 & \cellcolor{lightgray} 11.6 & 6.4 & - \\
& p-value          & .07 & .12 & .02 & - & - & .02 & \cellcolor{lightgray} $<.01$ & \cellcolor{lightgray} $<.01$ & .28 & - \\ \hline
\end{tabular}
\\ W: White; B: Black; P: Pardo; Y: Yellow; I: Indigenous; All: All ethnic groups compared simultaneously.
\label{tab:04}
\end{table}